\documentclass[fleqn,usenatbib]{mnras}
\usepackage[english]{babel}
\usepackage{graphicx}
\usepackage{hyperref}
\usepackage{natbib}
\usepackage{subfiles}
\usepackage{newtxtext,newtxmath}
\usepackage[T1]{fontenc}
\usepackage{ae,aecompl}
\usepackage{pdflscape}
\usepackage{amsmath}	
\usepackage{soul}

\title[A phylogenetic approach to galaxy evolution]{A phylogenetic analysis of galaxies in the Coma Cluster and the field: a new approach to galaxy evolution}

\author[Mart\'inez-Mar\'in et. al.]{
M. Mart\'inez-Mar\'in$^{1}$\thanks{E-mail: monserratlmartinezmarin@gmail.com},
R. Demarco$^{1}$,
G. Cabrera-Vives$^{2, 3}$,
P. Cerulo$^{2}$,
\newauthor { N. W. C. Leigh$^{1,4}$ and R. Herrera-Camus$^{1}$
}
\\
$^{1}$ Department of Astronomy, Facultad de Ciencias F\'isicas y Matem\'aticas, Universidad de Concepci\'on, Concepci\'on, Chile \\
$^{2}$ Department of Computer Science, Universidad de Concepci\'on, Concepci\'on, Chile \\
$^{3}$Millennium Institute of Astrophysics, Chile\\
$^{4}$Department of Astrophysics, American Museum of Natural History, New York, NY 10024, USA}

\date{Accepted October 2020. Received June 2020}

\pubyear{2020}

\hypersetup{draft}
\begin{document}
\label{firstpage}
\pagerange{\pageref{firstpage}--\pageref{lastpage}}
\maketitle

\begin{abstract}

We propose a phylogenetic approach (PA) as a novel and robust tool to detect galaxy populations (GPs) based on their chemical composition. The branches of the tree are interpreted as different GPs and the length between nodes as the internal chemical variation along a branch. We apply the PA using 30 abundance indices from the Sloan Digital Sky Survey to 475 galaxies in the Coma Cluster and 438 galaxies in the field. We find that a dense environment, such as Coma, shows several GPs, which indicates that the environment is promoting galaxy evolution. Each population shares common properties that can be identified in colour–magnitude space, in addition to minor structures inside the red sequence. The field is more homogeneous, presenting one main GP. We also apply a principal component analysis (PCA) to both samples, and find that the PCA does not have the same power in identifying GPs.

\end{abstract}


\section{Introduction}

In recent years, phyloinformatic studies have been used in astronomy to help understand different processes that could involve evolution. These studies were based either on discrete character states (DCs) or distance matrices of pairwise dissimilarities (Dm). Their main objective is to study the evolutionary history and relationships among individual or groups of astronomical objects. Astrocladistic utilizes DC, by either qualitative (e.g. morphology) or quantitative (e.g. luminosity, stellar mass, chemical composition) properties of galaxies to compare states of evolution in a cladogram. Astrocladistics has been used to study different astronomical objects like dwarf galaxies of the Local Group (\citealt{Fraix_Burnet_2006}), globular clusters and their classification (\citealt{Fraix_Burnet_2006}), and stellar populations (SPs) in $\omega$ Centauri (\citealt{Fraix_Burnet_2015}).

On the other hand, studies based on Dm are proper of Phylogenetics. This method takes as input the abundances of chemical elements of astronomical objects as a proxy for DNA in order to analyze their evolutionary history through their chemical compositions. This generates a branching diagram known as a phylogenetic tree. The branches of this tree are interpreted as different populations, and the position on the tree as a parameter to measure the mean rate of chemical enrichment of the population. This method has been employed as an attempt to reconstruct the chemical history of stars in the solar neighbourhood (\citealt{Jofre_2017}) and to retrieve stars from an open cluster (\citealt{Blanco_Cuaresma_2018}). 

Galaxies are complex systems made up of stars, gas, dust and dark matter. Understanding their formation and evolution is one of the key problems in modern astrophysics. Galaxy evolution is the result of several complex physical processes that involve the various galaxy components in different ways through cosmic epochs. These mechanisms bring galaxies to transition from one physical state to another and to change their observed properties.

Morphology is the most immediate way to classify galaxies since it is based on a visual assessment. Following \cite{Hubble_1926}, galaxies are classified as ellipticals, lenticulars or S0s, and spirals. A class of irregular galaxies is added to take into account objects that cannot be classified either as elliptical, S0 or spiral. Ellipticals and S0s are commonly said early-type galaxies (ETGs), while spirals and irregulars are commonly grouped together as late-type galaxies (LTGs). 

Morphology is related to the underlying structural distribution of the stellar populations and interstellar medium. ETGs tend to be red and with no or little star-formation, while LTGs tend to be blue and star-forming (see \citealt{Baldry_2004}, \citealt{cassata08}, \citealt{taylor15}). Up to $z \sim 6$, the star formation in galaxies is observed to be related to their stellar mass: massive galaxies are less star-forming than their low-mass counterparts \citep{Cowie_1996}. The stellar mass of galaxies is finally related to their metallicity: more massive galaxies are also metal-richer (\citealt{1979A&A....80..155L}; \citealt{2004ApJ...613..898T}; \citealt{2005MNRAS.362...41G}) 

Regardless of the environment where galaxies reside, they show a bimodal colour distribution. Three main populations can be defined from this color-magnitude distribution: (1) a Red Sequence of quiescent galaxies; (2) a Green Valley as an "intermediate" population; and (3) a diffuse cloud of blue star-forming galaxies. Thus, in principle, the evolution of a galaxy population can be investigated by looking at the gradual build-up of the Red Sequence as a function of redshift (\citealt{Tanaka_2010}; \citealt{Gobat_2011}; \citealt{Muzzin_2012}; \citealt{Lidman_2004}; \citealt{Snyder_2012}). 

Nevertheless, we can obtain more detailed information by studying galaxy spectra.
Spectra of galaxies contain information about the properties of both gas and stellar populations.  These contain a preserved record of the host galaxy's formation and evolution. Emission lines are used to study the physical state of the ionized gas, so we can trace black hole accretion, derive gas kinematics and star formation activity (\citealt{Kauffmann_04}; \citealt{Schawinski_07}). Absorption lines provide information about the properties of stellar populations and can be used to derive ages, metallicities, star formation histories and elemental abundances.(\citealt{DRSEPR93}; \citealt{FISHER96}; \citealt{Jorgensen_99}; \citealt{Kuntschner_00}; \citealt{Thomas_10}). The most common sets of optical absorption line indices are the Lick index system (\citealt{Burstein_1984}; \citealt{Faber_1985}; \citealt{Worthey_1994}; \citealt{Worthey_1997}), and the [OIII] indices (\citealt{Gonzz1993}). These indices group different absorption lines together.  Therefore, they measure the compositions of various chemical elements.  However, the abundance derivations for individual elements is non-trivial.

Studies use absorption lines to compare morphology with chemical abundances and other properties of the host galaxy such as velocity dispersion (\citealt{Davies_87}, \citealt{Burstein_1998}), B-band absolute magnitude ( \citealt{Faber_1973}; \citealt{Terlevich_1981}), or effective radius (\citealt{Parikh_2018}). 

In this work, we propose a new technique to study galaxy evolution, analyzing galaxies according to their chemical composition through a phylogenetic approach. This approach generates a phylogenetic tree that gives a graphical visualization of the chemical diversification between galaxies through a branching process. Each branch gathers galaxies with similar chemical composition, and their hierarchical structure can be related to common properties of galaxies like stellar mass ($M_*$), specific star formation rate (sSFR; $SFR/M_*$), metallicity and morphology. The main characteristic of this schema is that it allows us to hypothesise how developmental changes are mediated by tracing branching points to a  physical process or galaxy property. This is critical for understanding the mechanisms that take part in the evolution of galaxies and, overall, finding common ancestors within a group of chemically related galaxies allows us to think about galaxy evolution and their possible morphological change as a non-linear process.

For this, we analyze the chemical composition of stellar populations in galaxies that reside in different environments. We consider 475 galaxies in the Coma cluster with redshifts in the range $0.015<z<0.033$, and 438 field galaxies at $0.035<z<0.054$. We use a set of 30 abundance indices from the value added catalogs of the Sloan Digital Sky Survey data release 14 (SDSS DR14) (\citealt{SDSSDR14}). We generate a phylogenetic tree based on a distance matrix from the pairwise differences of these indices using the Neighbor Joining algorithm (NJ; \citealt{Saitou_1987}; \citealt{Studier_1988}). NJ allows us to find hierarchical relationships that can be used to uncover possible evolutionary paths of the observed galaxies. We interpret the branches of the phylogenetic tree as different galaxy populations (GP), and the length between the nodes of these branches as a gradient of chemical difference in the stellar population content of these galaxies. In this way we can see how properties like morphology, stellar mass, sSFR or metallicity ([Z/H]) relate with the galaxy populations found with this phylogenetic approach at different environments. 

\section{Data}
Previous phyloinformatic studies in astronomy, have utilized abundance ratios of several elements as a proxy for DNA. In our case study, we will use instead absorption line indices, which measure different individual element abundances giving us information about the stellar population properties of our galaxies. On the other hand, in order to understand which physical properties of galaxies are involved in the hierarchical structure of the phylogenetic trees, we will use information about their stellar mass, specific star formation rate (sSFR), metallicity (Z/H) and stellar ages. 

\subsection{Phylogenetic Study}

For the phylogentic study on the Coma cluster we use the membership of \citealt{Beijersbergen_2002}, obtaining 475 galaxies with redshifts between $0.015$ and $0.033$. For the field, we select galaxies that do not belong to galaxy groups (\citealt{Tempel_2012}) nor filaments (\citealt{Tempel_2014}), however the galaxies that meet this criterion and have information of abundance indices are too small to use as a sample comparable to that of the Coma cluster. Therefore, we select a sample of galaxies that meets the above criteria but at a slightly larger redshift. Thus obtaining a sample of 438 galaxies with redshifts between $0.035$ to $0.054$. For both samples we obtain the spectroscopic data from the seventh data release (DR7 \citealt{SDSSDR7}) of the Sloan Digital Sky Survey (SDSS;\citealt{York_2000} ). We use the spectral line strength measurement database provided by the OSSY group (\citealt{Oh_2011}), considering the stellar absorption-line measurements for this galaxy sample. 

\subsection{Ancillary Data}  

The stellar mass is based on fits to the photometry following (\citealt{Kauffmann_04}), and (\citealt{Salim_2007}) from the SDSS DR7.
 The galaxy sample of the Coma cluster presents a stellar mass distribution characterized by a mean value of $\log(M_{*}/M_{\odot}) = 9.98$ and a 1-$\sigma$ dispersion of $\log(M_{*}/M_{\odot}) = 0.58$. For the field sample, we have a stellar mass distribution centered at $\log(M_*/M_\odot) = 9.99$ with a 1-$\sigma$ dispersion of $log(M_*/M_{\odot}) = 0.50$.
The $sSFR$ estimates are based on the technique discussed in \citealt{Brinchmann_2004}, with a slight modification for non-star-forming galaxies. Here, the likelihood distribution of the sSFR is constructed as a function of D4000 using the star-forming sample, also from the SDSS DR7. Our sample of cluster galaxies in Coma presents a sSFR distribution centered at $\log{(sSFR/yr^{-1})} = -10.28$ with a 1-$\sigma$ dispersion $\log{(sSFR/yr^{-1})} =0.53$, on the other hand, the Field presents a sSFR  distribution centered at $\log{(sSFR/yr^{-1})} = - 9.39$ with a 1-$\sigma$ dispersion of $\log{(sSFR/yr^{-1})} = 0.6$) (see Figure \ref{fig2}). 
 We calculate the ages and metallicities of the galaxies in the GPs found by the PA following the procedure from \citealt{Cardiel_2003}. We interpolate our model with the simple stellar population predictions of \citealt{Thomas_2010} for the $H_{\beta}$ and $C4668$ indices. 

Finally, we utilize the visual morphological classification from \cite{Beijersbergen_2002} and two parameters of the morphological classification from \citealt{Dom_2018}, obtained with deep learning algorithms implementing convolutional neural networks (CNNs).  \citealt{Dom_2018} used T Types (see \citealt{DeVauc_1959}) to separate ETGs from LTGs and gives the probability $P_{S0}$ of being $S0$ versus a pure elliptical in order to have three galaxy classes: Elliptical, Spiral and S0. Elliptical galaxies are defined such that they have TType values ranging from $-3$ to $0$ and $P_{S0}$ values lower than $0.5$. S0 galaxies have TType values between $-1$ and $3$, and $P_{S0}$ values higher than $0.5$. Finally, spiral galaxies have a TType value between $3$ and $8$, independent of their $P_S0$ value. For galaxies in the sample that do not have information on TType and $P_{S0}$, the visual morphological classification from \cite{Beijersbergen_2002} is assigned. 

\section{The method}
\label{method} 

We perform a phylogenetic study based on a distance matrix with the NJ algorithm, utilizing the open-source Scikit-bio python package. In our case study, the distance matrix elements are obtained by the summation of chemical distance differences for a given abundance index ($X_{k}$) ($k$ is the specific absorption index) between every pair of galaxies ($i$ and $j$) in our sample. Therefore, each element of our distance matrix is the total chemical distance $D_{i,j}$ from our set of abundance indices between our galaxies:
 
\begin{equation*}
D_{i,j}=\sum^{N}_{k=1}|[X_{k}]_{i}-[X_{k}]_{j}|.
\end{equation*}
Where N is the number of end nodes.
For example, let us consider the following distance matrix:

\[
\begin{bmatrix}
         & G_{A} & G_{B} & G_{C} & G_{D} & G_{E}& G_{F}\\
G_{A}    &  0    &   5   &  4    &  7    &  6   & 8   \\
G_{B}    &  5    &   0   &  7    &  10   &  9   & 11  \\
G_{C}    &  4    &   7   &  0    &  7    &  6   & 8   \\
G_{D}    &  7    &   10  &  7    &  0    &  5   & 9   \\
G_{E}    &  6    &   9   &  6    &  5    &  0   & 8    \\
G_{F}    &  8    &  11   &  8    &  9    &  8   & 0   \\
\end{bmatrix}
\]
A clustering method would group galaxies A and C as the most similar galaxies, because their total chemical distance $D_{i,j}$ is the smallest. This assume that the elements under study evolve in a clock-like behavior. Therefore, the samples obtained at the same calendar time would have experienced the same number of generations since their common ancestor, maintaining a constant rate of change per generation. Nevertheless, in galaxies there are several internal and external processes, especially in large and dense cosmic structures where they may undergo interactions with the surrounding environment that make them deviate their course from a linear evolution. The main characteristic of NJ is that it takes into account the development of evolution at different rates.
For that, the NJ defines a new corrected distance $D^{'}_{i,j}$ which subtracts the
divergence $r_i$ and $r_{j}$. 

\begin{equation*}
  D^{'}_{i,j} =  D_{i,j} - (r_{i} + r_{j})  
\end{equation*}

The divergences, correspond to the mean over the total chemical distance $D_{i,j}$ of each possible pair of galaxies and the other galaxies:

\begin{equation*}
    r_{j}=\frac{\sum_{i\leq j}D_{i,j}}{N-2}.
\end{equation*}

\begin{figure}
    \centering
    \includegraphics[width=\columnwidth]{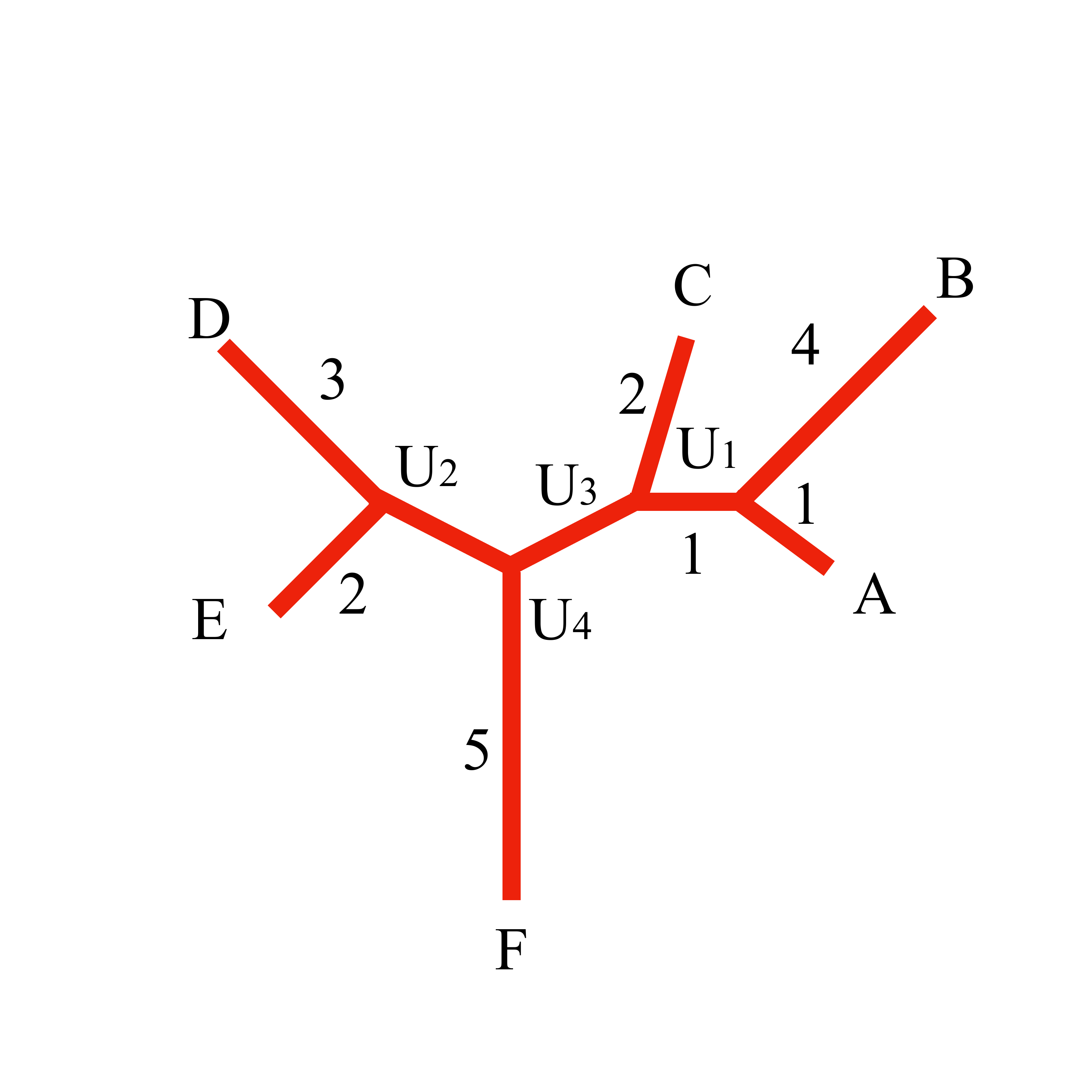}
    \caption{Final star representation of the galaxy distribution in our example. Red branches represent known distances. Each galaxy is at the end of a respective branch. The U letters represent the nodes that join our galaxies. Numbers indicate the chemical distance between galaxies and nodes or between nodes. We have a topology that shows two different galaxy populations: A, B and C versus E and D. Galaxy F does not belong to any galaxy population. }
    \label{fig1}
\end{figure}

With this new rate-corrected distance, we create a new distance matrix that will show us how to obtain our first node $U_{1}$ that joins the most similar galaxies, which in our example are galaxies A and E, for which $D^{'}_{i,j}$ is minimal. However, we could also choose galaxies D and E. The decision of selecting one or the other node will define the zero point of our tree. We can calculate the length of the branches from each galaxy to this node:

\begin{equation*}
    S_{i,j}=\frac{D_{i,j}+(r_{i}-r_{j})}{N-2} \cdot \frac{1}{2},
\end{equation*}
where $N$ is the number of end-nodes. In order to get the fully resolved topology of our tree, we now repeat the process until we have no galaxies left (see Figure \ref{fig1}). 

With this algorithm galaxies that have similar chemical composition, such as A and B, belong to the same node at the top of the tree. Their next closest neighbour in stellar content is galaxy C. Galaxies A, B and C belong to the same branch which we interpret as the same GP. In contrast, galaxies E and D belong to another branch separated by node $U_{4}$ and, therefore, to a different GP. On the other hand, galaxy F does not belong to any GP.
The distance between nodes $U_1$ and $U_3$ works as a chemical index that will allow us to differentiate galaxies A and B from galaxy C, even when they belong to the same population.
Finally, galaxies that are in the tree canopy are the galaxies whose chemical composition are most similar, and, therefore, they have the longest distance to the root.
Galaxies at the top of the tree end up being actually chemically different from galaxies at the root of the tree, even when they belong to the same population. Therefore the length between the nodes (Hereafter: NodeLength) along the tree, will help us understand which parameters can promote the chemical diversification of galaxies from the top of the tree down to the root. 

In order to obtain a consistent hierarchical structure for the phylogenetic tree, we calculate 1,000 trees utilizing Monte Carlo sampling from a normal distribution where the center of the distribution is the value of the absorption index and the width the error of the measurement from SDSS DR14 (\citealt{SDSSDR14}). Also, we make bootstrapping with replacement so that every time we calculate our distance matrix with Monte Carlo sampling we will have a different combination of 29 or 30 absorption indices to generate our tree. Finally, we make a majority rule consensus, which considers structures in the tree that repeat at least $50\%$ of the time in the trees that we calculate. This procedure allows us to eliminate branches that are not statistically significant.

\section{Analysis}

\begin{table}
 \caption{Galaxy distribution of different structures in the consensus phylogenetic tree of the Coma Cluster}
 \begin{tabular}{ll}
  \hline
    Population & No. of members   \\
  \hline
  \verb'Branch 1' & 10\\[2pt] 
  \verb'Branch 2' &18 \\[2pt]
  \verb'Branch 3'  & 67\\[2pt]
  \verb'Branch with 1 node'  & 90\\[2pt]
  \verb'Branch with 2 node'  &39 \\[2pt]
  \verb'Branch with 3 node' &18\\[2pt]
  \verb'Branch without nodes' &233\\[2pt]
  \hline
\end{tabular}
\label{table:1}
\end{table}

\subsection{The Coma Cluster} 
\label{Coma} 
We calculate a phylogenetic tree for our galaxy sample of the Coma cluster. Our tree presents three main branches that we arbitrarily named branch one (B1), branch two (B2) and branch three (B3). From the majority consensus rule of 1,000 trees, they appear in 683, 694 and 655 trees, respectively. Several other minor branches are found that appear at least 50$\%$ of the times and whose complete description is shown in Table \ref{table:1}.

Each of our branches represents a GP and they are differentiated by their chemical content. In what follows, we will focus on the three main GPs, namely B1, B2 and B3. In order to prove that our GPs are different, we plot them in Color-Magnitude space, indicating also galaxy morphology.

\begin{figure}
 \includegraphics[width=\columnwidth]{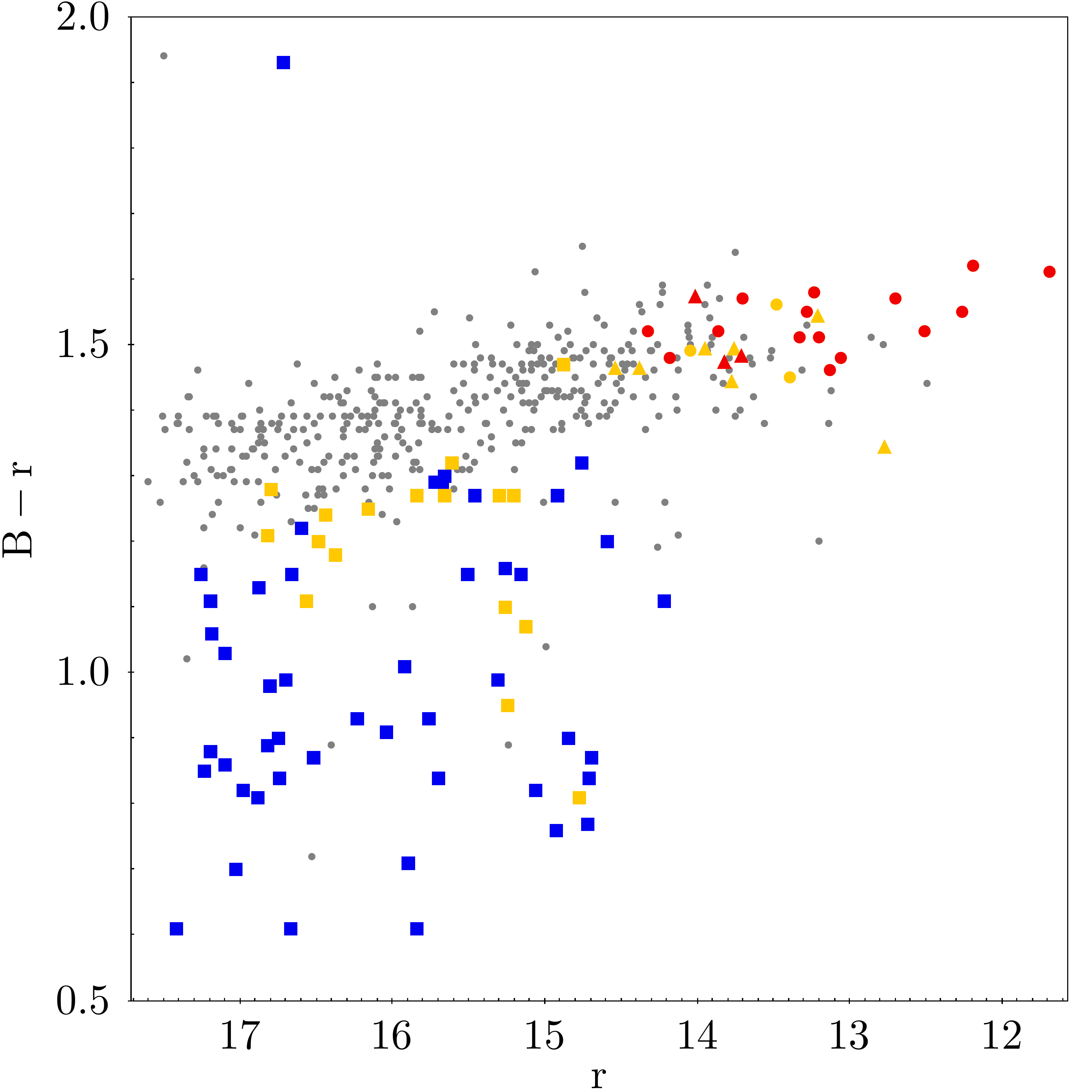}
 \caption{Color-Magnitude diagram of the Coma Cluster with photometric filters Johnson B and Sloan r. Triangles, circles and squares represent galaxies belonging to branch 1 (B1), branch 2 (B2) and branch 3 (B3), respectively. The colors indicate galaxy morphology where red corresponds to elliptical galaxies, blue to spiral galaxies, and yellow to S0 galaxies. Grey dots represent galaxies from the Coma cluster that not appear in the main branches of the consensus phylogenetic tree.}
 \label{fig2}
\end{figure} 

\begin{figure}
 \includegraphics[width=\columnwidth]{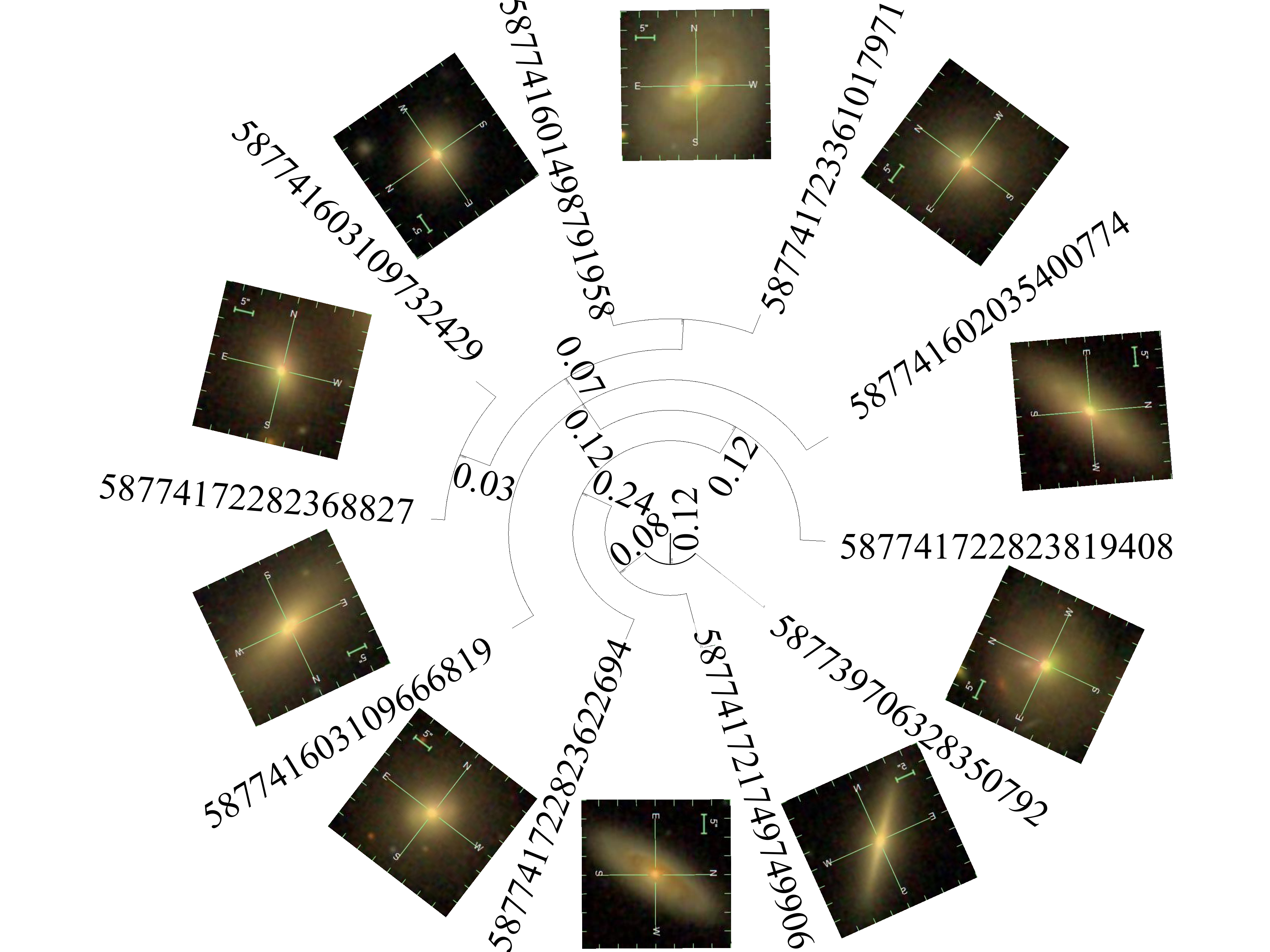}
 \caption{Branch B1 of the phylogenetic tree of the Coma Cluster that represents our first galaxy population. Each galaxy has its ID number from the SDSS DR7, and its image from the SkyServer DR15. The number between the nodes is the chemical length of their separation.}
 \label{fig3}
\end{figure}

\begin{figure*}
 \includegraphics[width=\columnwidth]{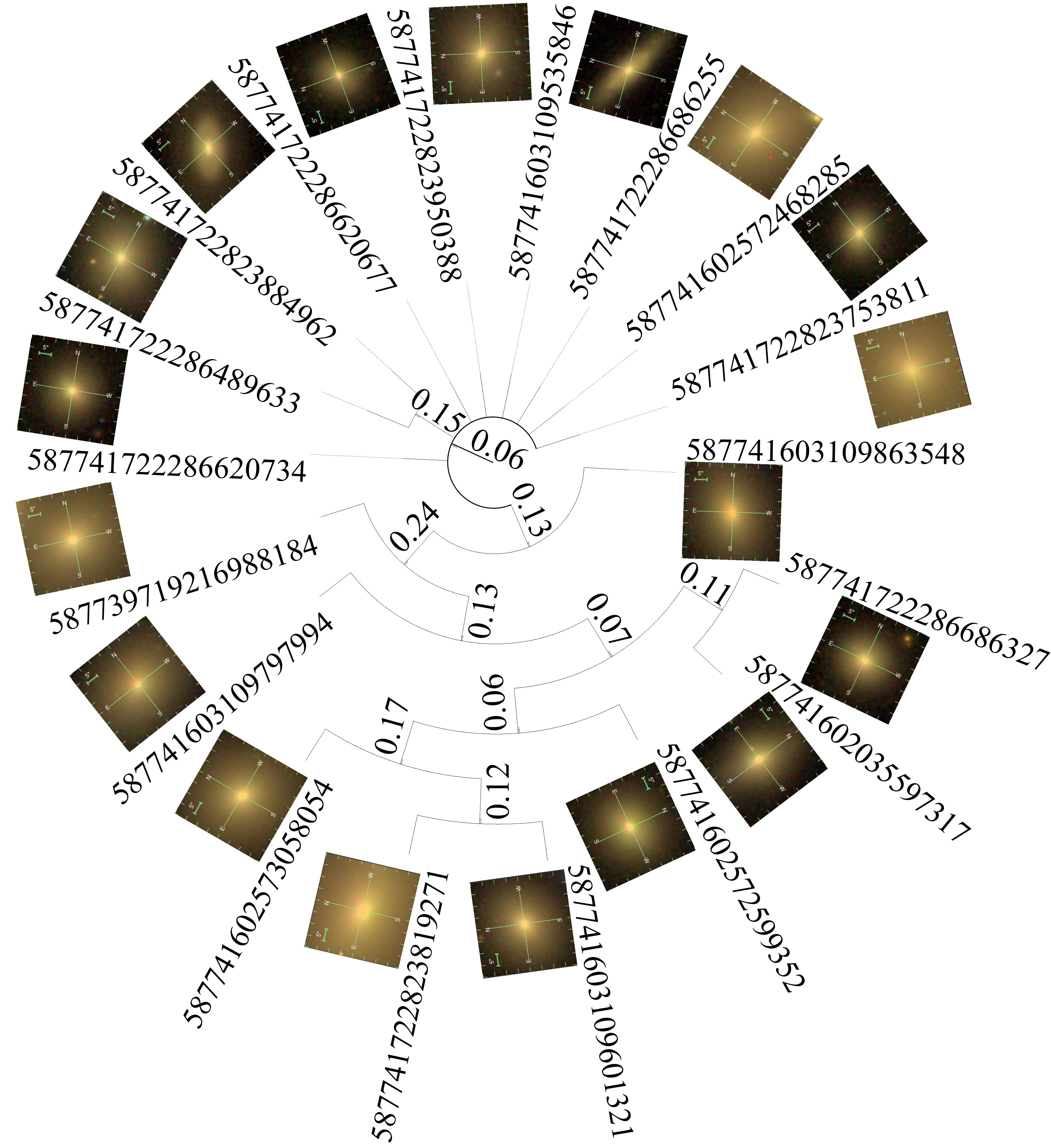}
 \caption{Branch B2 of the phylogenetic tree of the Coma Cluster that represents our second galaxy population. Each galaxy has its ID number from the SDSS DR7, and its image from the SkyServer DR15. The number between the nodes is the chemical length of their separation.}
 \label{fig4}
\end{figure*}

\begin{figure*}
 \includegraphics[width=17cm]{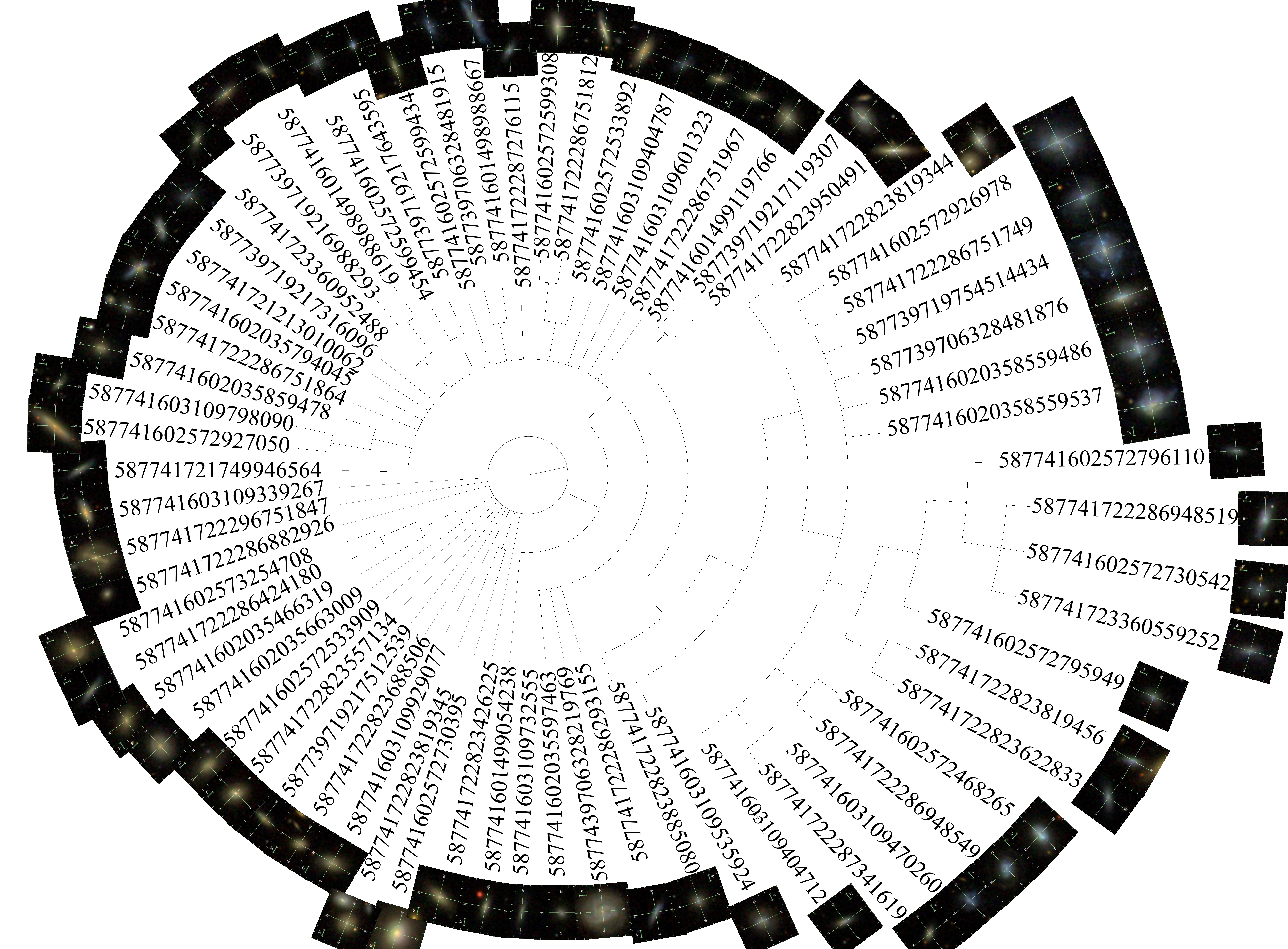}
 \caption{Branch B3 of the phylogenetic tree of the Coma Cluster that represents our third galaxy population. Each galaxy has its ID number from the SDSS DR7, and its image from the SkyServer DR15. The number between the nodes is the chemical length of their separation.}
 \label{fig5}
\end{figure*}

Figure \ref{fig2} shows that, branches B1 and B2 belong to the bright-end of the Red Sequence, and B3 belongs to the blue cloud of the cluster. This shows us that at least populations B1 and B2 are different from population B3, as we see they belong to different areas in the Color-Magnitude diagram. From the consensus morphology, we see that B1 has 3 elliptical galaxies and 7 S0 galaxies; B2 has 15 elliptical galaxies and 3 S0 galaxies; whereas B3 has 50 spiral galaxies and 17 S0 galaxies. In order to have a better visualization of this GP'S, we plot each population as a single tree with an image of they morphology. (See Figures \ref{fig3}, \ref{fig4} and \ref{fig5}). 

\begin{figure}
 \includegraphics[width=\columnwidth]{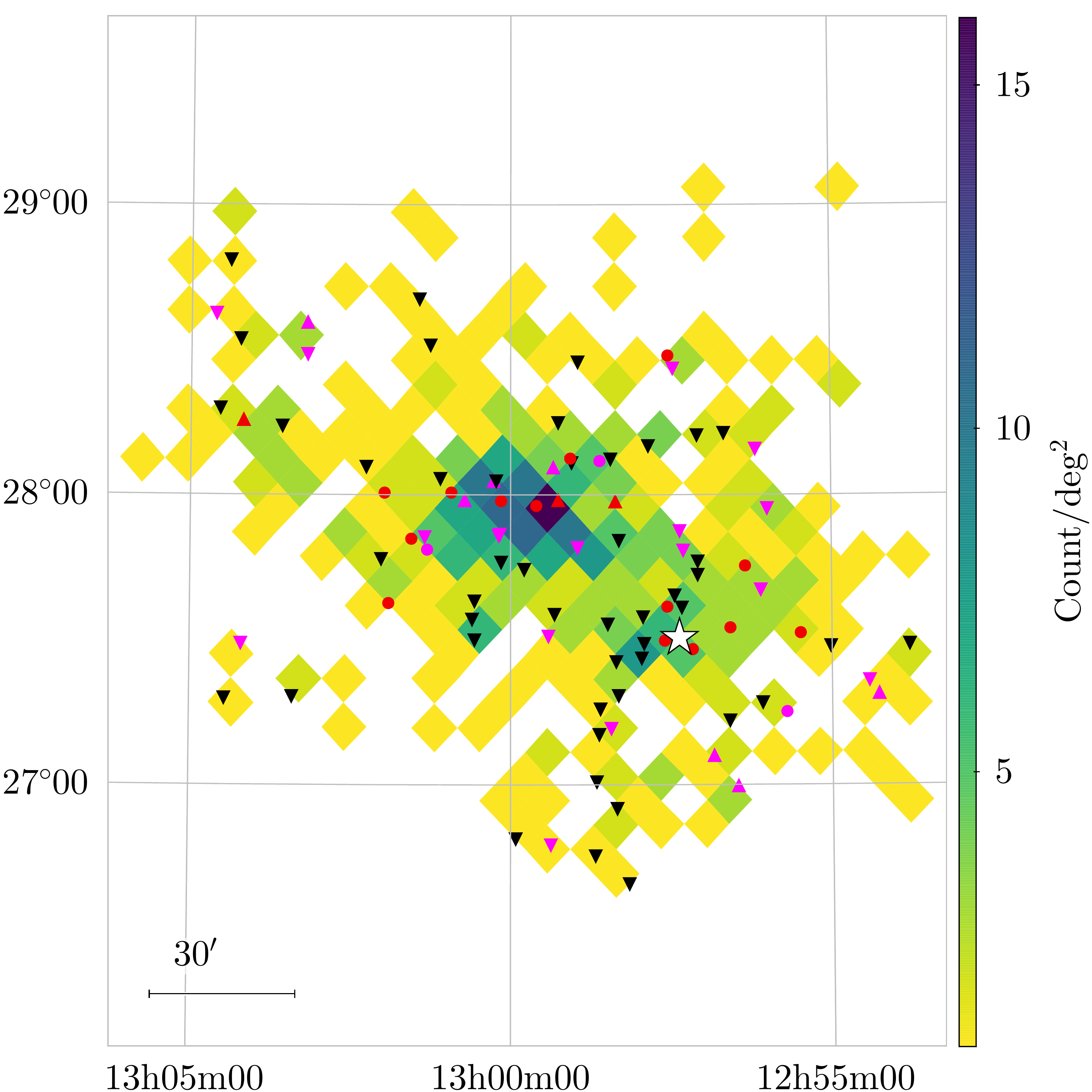}
 \caption{Projected sky distribution of galaxies in the Coma Cluster belonging to different branches of the phylogenetic tree. Triangles, circles and inverted triangles represent galaxies belonging to branch 1 (B1), branch 2 (B2) and branch 3 (B3), respectively. The colors indicate the morphology of the galaxy where red corresponds to elliptical galaxies, black to spiral galaxies, and magenta to S0 galaxies. The color scale shows the number count of galaxies by degree squared in the sky for the Coma Cluster. The white star points out the galaxy NGC4839.}
 \label{fig6}
\end{figure}

We plot our GP's in a sky projection of the galaxy cluster in Figure \ref{fig6}. This shows us that, 1) galaxies in the B1 population have lenticular morphology at the outskirts of the cluster and elliptical morphology in the cluster core; 2) galaxies in the B2 population are distributed in two clumps in the cluster, one in the centre and the other near the galaxy NCG 4839 which is associated with a subgroup of galaxies in-falling into the Coma Cluster. This population includes the two brightest cluster galaxies (BCG) NGC4874 and NGC4889 of the Coma cluster (\citealt{Neumann_2001}); and 3) galaxies in the B3 population do not show any preferential distribution in the cluster. 

\begin{figure}
 \includegraphics[width=\columnwidth]{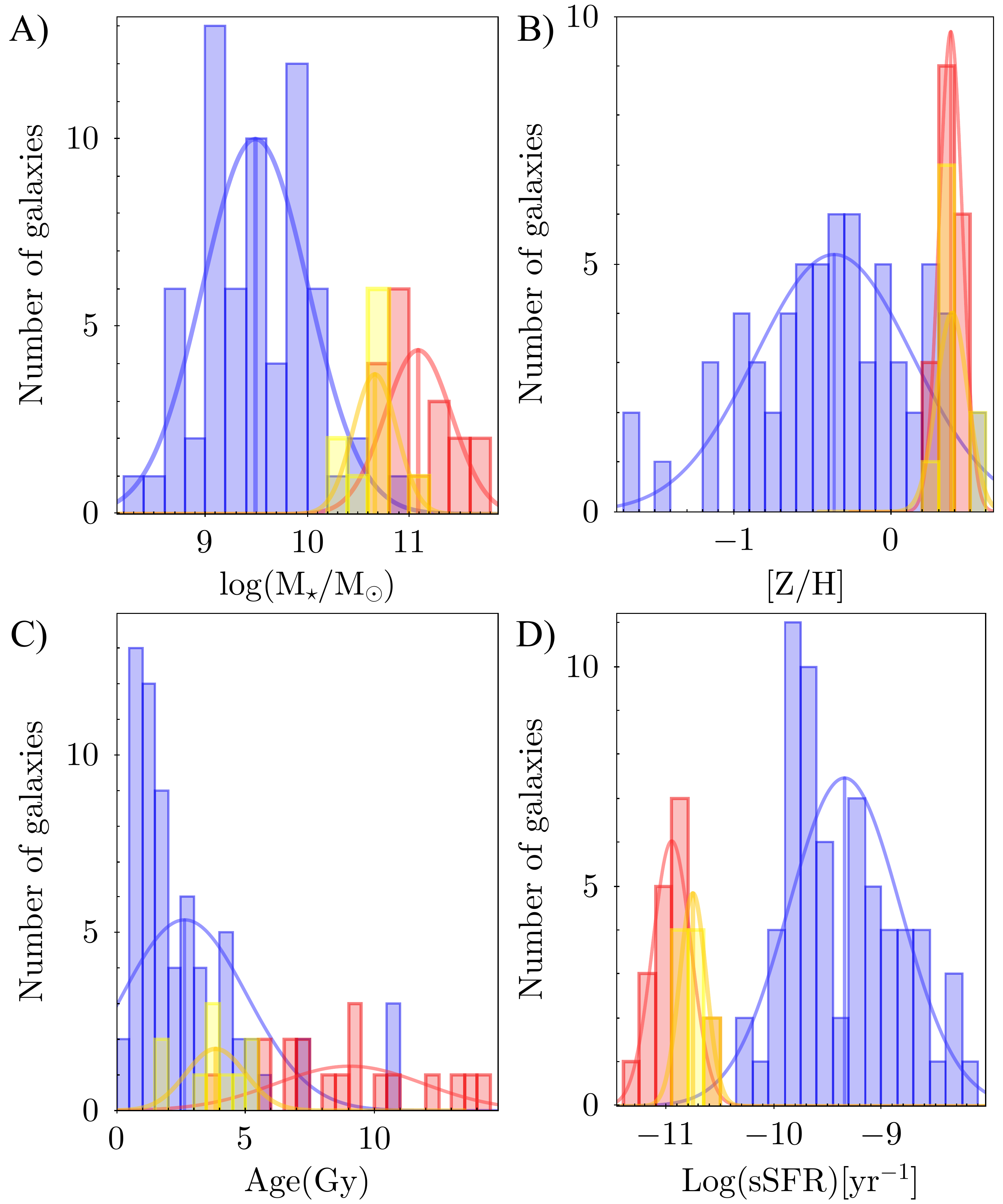}
 \caption{The histograms A, B, C and D show the distribution of galaxies from the three different branches (B1, B2 and B3) with respect to stellar mass, metallicity, age and sSFR. The yellow, red and blue colors represent galaxies from branches B1, B2 and B3, respectively. Each branch sample have a gaussian function, where the mean is indicated with a vertical line of the corresponded color sample.}
 \label{fig7}
\end{figure}

In order to characterize the common properties of the GP's, we compare their stellar masses, sSFRS, metallicities and ages in Figure \ref{fig7}. B1 galaxies are quiescent with a high stellar mass with a mean value of $log(M/M_{\odot}) = 10.66$ and 1 $\sigma$ dispersion of $ log(M/M_{\odot}) = 0.21 $, high metallicity with a distribution centered at $[Z/H] = 0.38$ and $1-\sigma$ dispersion of $ [Z/H] = 0.1)$, low sSFR with a mean value of $10^{-10.75} [yr^{-1}]$ and 1-$\sigma$ dispersion of $0.13 yr^{-1}$, but of predominantly S0 morphology and with a distribution of young-to-intermediate age stellar populations centered at $3.82 Gyr$ and 1-$\sigma$ dispersion of $1.15 Gyr$. On the other hand, B2 galaxies correspond to quiescent galaxies with high stellar mass with a mean value of $log(M/M_{\odot}) = 11.08$ and $1-\sigma$ dispersion of  $log(M/M_{\odot}) = 0.33$, high metallicity with a distribution centered at $[Z/H] = 0.37$ and 1-$\sigma$ dispersion of $ [Z/H] = 0.07$), low sSFR with a mean value of $10^{-10.95} yr^{-1}$ and 1 $\sigma$ dispersion of $0.18 yr^{-1}$, intermediate-to-old-age stellar populations with a distribution centered at $ 9.0 Gyr $  with 1 $\sigma$ dispersion of  $2.85 Gyr$ and mostly elliptical morphology. B3 galaxies have low stellar mass with a mean value of $log(M/M_{\odot}) = 9.47$ and 1-$\sigma$ dispersion of $log(M/M_{\odot}) = 0.51$, with low metallicity distribution centered at $[Z/H] = -0.33$ and 1-$\sigma$ dispersion of $log(M/M_{\odot}) = 0.49$ and high sSFR centered at $10^{-9.64} yr^{-1}$ with 1-$\sigma$ dispersion of $1.8 yr^{-1}$, have predominantly spiral morphology and contain stellar populations having a wide range of ages $2.62 Gyr$ with 1-$\sigma$ dispersion of $ 2.41 Gyr$). From this comparison we see that although branches B1 and B2 reside in the red sequence of the cluster, they have different distributions of ages and mass. Therefore from this comparison we can conclude that B1, B2 and B3 are indeed three different GPs. 

Now we analyze each branch's hierarchical structure. 
As we show in section \ref{method}, the NodeLength works as a chemical index. This chemical index characterizes the variation in the chemical composition of galaxies. In order to see how the general properties of mass, sSFR, ages and $[Z/H]$ relate internally in our GP, we add the node length from the root of the tree to the leaves where each galaxy is. 

\begin{table}
 \caption{Statistical Results between Nodelenght and galaxy properties for the Coma cluster main populations. We show the slope and y-intercept, with the 1-$\sigma$ uncertainties for a OLS fitting, as well as the 1-$\sigma$ dispersion for the WLS and the value for the chi squared forced to unity.}
 \begin{tabular}{lllllll}
  \hline
   Branch & Slope & $y_{intercept}$ & $\overline{X}_{OLS}$ & $\sigma_{OLS}$ & $\sigma_{WLS}$ & $X_{i}^{2}$  \\
  \hline
   \hline
\verb"Mass"&&&&&&\\[2pt] 
 \hline
  \verb'B1' & 0.14  &10.58 & 10.66 & 0.04& 3.62&22.68\\[2pt] 
  \verb'B2' & 0.31 &10.95 & 11.08&0.11 &5.56 &50.48\\[2pt]
  \verb'B3' &-0.16  &9.63 & 9.47&0.13&4.91 &36.73\\[2pt]
   \hline
\verb"[Z/H]"&&&&&&\\[2pt] 
 \hline
  \verb'B1' & 0.25 &0.23  & 0.39&0.07 &0.17 &0.01 \\[2pt]
  \verb'B2' & 0.06 &-0.39 & 0.37& 0.02&0.23&0.07 \\[2pt]
  \verb'B3' & -0.17 & -0.21 &-0.37 & 0.13 & 0.29& 0.03\\[2pt]
   \hline
\verb"Age"&&&&&&\\[2pt] 
 \hline
  \verb'B1' & -3.58& 6.03& 3.83&1.06 & 1.0 &7.32 \\[2pt] 
  \verb'B2' &  3.65 &7.48 & 9.02&1.31 & 5.56& 24.33\\[2pt]
  \verb'B3' &1.38 &1.21 &2.55 &1.08 &2.68 & 3.17\\[2pt]
  \hline
\verb "sSFR" &&&&&&\\[2pt]
\hline
\verb'B1' & -0.22 & -10.62& -10.75 & 0.06 & 4.73 & 22.62\\[2pt] 
\verb'B2H' & 0.06  & -10.98& -10.95 & 0.02 & 5.06 & 51.02\\[2pt]
\verb'B3' & 0.32  &-9.67  & -9.36  & 0.25 & 4.74 & 36.9 \\[2pt]
\hline 
\end{tabular}
\label{table:2}
\end{table}
 We investigated the variations in stellar mass, metallicity,
stellar age and sSFR as a function of NodeLength to
establish whether one of these physical properties is
driving the differences in chemical composition along a
branch. For that we use a bootstrap methodology generating 1,000 fake data sets with replacement to sample stellar mass, sSFR, metallicity and stellar age distributions. Then we use an ordinary least
squares regression (OLS) to get the slope and y-intercept in
order to report the standard deviation of the distribution. We also perform a weighted least squares (WLS) in order to assess the error in our absorption indices, setting the uncertainties equal to some intrinsic dispersion term $\sigma_{int}$ and the evaluating the value of this $\sigma_{int}$ term by forcing the reduced chi-squared of your fit to be unity. We report the value that gives $X_{i}^{2} = 1$, and the $\sigma_{int}$ for the WLS fit in Table \ref{table:2}

 We find from Table \ref{table:2} that our Nodelength parameter have a good fit with the metallicities of our populations. Therefore the metallicity plays an important roll in the formation of the hierarchical structures of the different branches of our phylogenetic tree. Population B3 has different metallicity content than populations B1 and B2, while populations B3 and B1 share similar ages, although stellar mass and sSFR are not relevant for their hierarchical relationships.

\begin{figure*}
 \includegraphics[width=18cm, height=22cm]{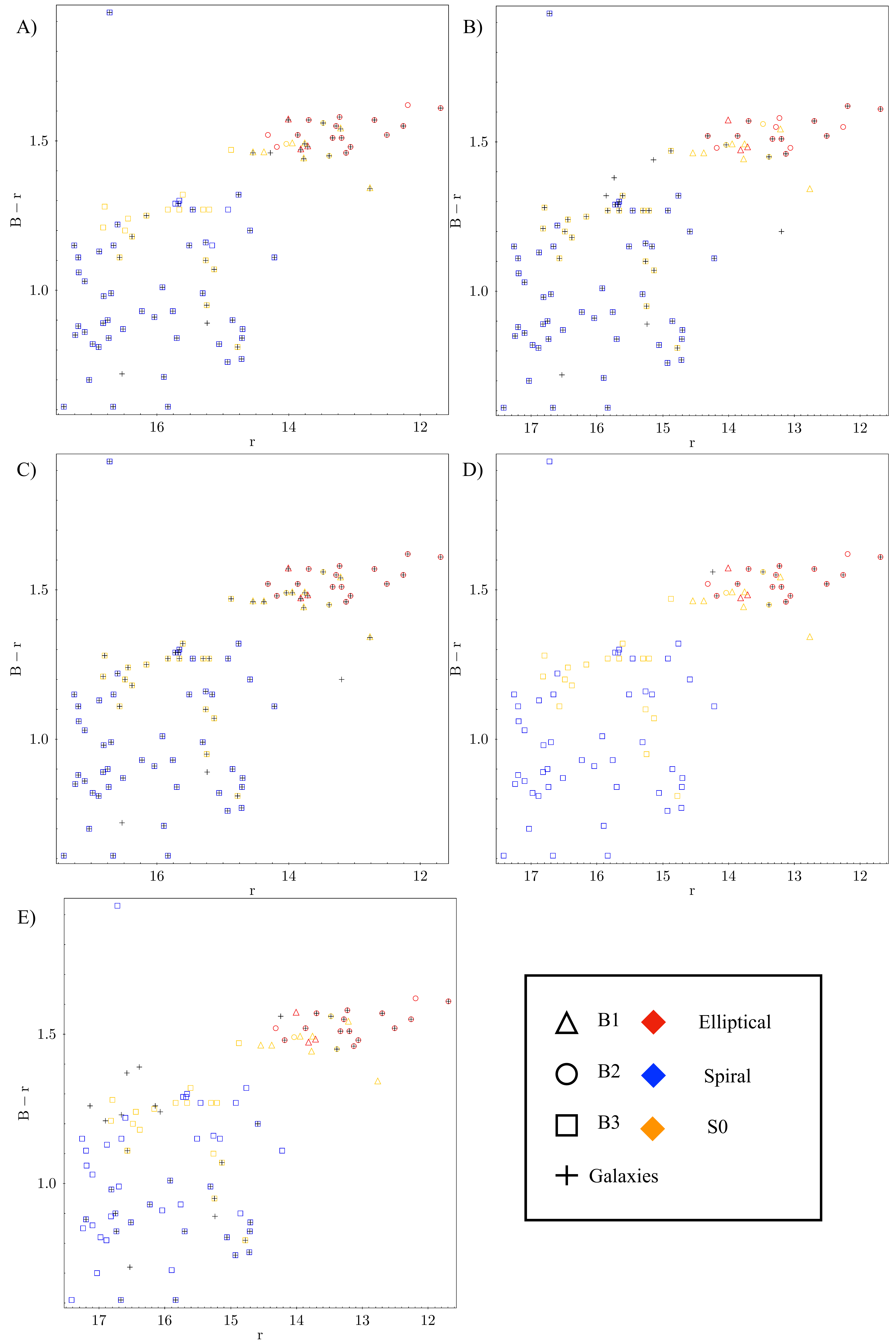}
 \caption{Color-magnitude diagram of our Coma Cluster galaxy sample with the photometric filters Johnson B and Sloan r. Panels A, B, C, D, and E show a recalculated phylogenetic tree without a negative response to $\alpha/Fe$ enhancement, insensitive response to $\alpha/Fe$ enhancement, positive response to $\alpha/Fe$ enhancement, absorption Balmer lines, and Emission Line $[OIII]\alpha 5007$ indices $[OIII]$. The black plus represent the galaxies that appear in the branches of each recalculated tree.}
 \label{fig8}
\end{figure*} 

\begin{table}
 \caption{Remaining galaxy populations for the recalculated phylogenetic trees without a specific set of line indices }
 \begin{tabular}{ll}
  \hline
   Index Class & Branches  \\
  \hline
  \verb'Blamer Lines' & B2\\[2pt] 
  \verb'Positive response to alpha/Fe enhancement' &B3 B2 B1 \\[2pt]
  \verb'Negative response to alpha/Fe enhancement'  & BE B3 B1\\[2pt]
  \verb'Insensitive to alpha/Fe'  & B3 B2\\[2pt]
  \verb'Emission Line $[OIII] \alpha 5007$'  &B2 B3  \\[2pt]
  \hline
\end{tabular}
\label{table:3}
\end{table}

Studying general or average properties such as metallicity, sSFR, age and mass is not enough to fully understand what are the physical differences and/or common characteristics of galaxies in our GPs. These are what determine the hierarchical structures we observe. Therefore we separate our abundance indices into five classes according to  \citealt{Thomas_03}: (1) absorption, age-sensitive Balmer lines including $H\delta_A$, $H\delta_F$, $H\gamma_A$, $H\gamma_F$, $H\beta$ and $H\beta_G$; (2) positive responses to $\alpha/Fe$ enhancement, including $CN_1$, $CN_2$, $Mg_1$, $Mg_2$, $Mg_b$, $Ca4227$ and $G4300$; (3) negative responses to $\alpha/Fe$ enhancement, including $Fe4383$, $Fe4531$, $Fe4930$, $Fe5015$, $Fe5270$, $Fe5270S$, $Fe5335$, $Fe5406$ and $Fe5709$; (4) emission Line $[OIII] \alpha 5007$, with $[OIII_{1}]$ and $[OIII_{2}]$ (\citealt{Gonzalez_2005}); and (5) insensitive to $\alpha/Fe$, comprising $NaD$, $C4668$, $Fe5782$, $Ca4455$, $TiO_1$ and $TiO_2$ . 

Next, we recalculate the phylogenetic tree for the Coma cluster with the same sample of galaxies but without one of these classes of abundance lines. We obtain five new consensus trees, so we can analyze how these classes influence the presence of our GPs in the phylogenetic tree. 
For this we analyze our re-calculated consensus tree, by plotting color-magnitude diagrams. We graph the distribution of galaxies from our original tree, and that of the galaxies that appear in our re-calculated trees (see Figure \ref{fig8}). We summarize the results in Table \ref{table:3}.

\begin{table}
 \caption{ Statistical Results between Nodelenght and absorption indices for the Coma cluster main populations. We show the slope and y-intercept, with the 1-$\sigma$ uncertainties for a OLS fitting, as well as the 1-$\sigma$ dispersion for the WLS and the value for the chi squared forced to unity. }
 \begin{tabular}{|l l l l l l l|}
  \hline
    Branch & Slope & $y_{intercept}$ & $\overline{X}_{OLS}$ & $\sigma_{OLS}$ & $\sigma_{WLS}$ & $X_{i}^{2}$  \\
 \hline
  \hline
$H_{\delta}A$&&&&&&\\[2pt] 
 \hline
  \verb'B1' & 1.3   & -1.39& -0.59 & 0.37& 0.11 &0.35\\[2pt] 
  \verb'B2' & -0.39 & -2.2 & -2.36 & 0.14& 1.51 &2.23\\[2pt]
  \verb'B3' & 1.56  &3. 35 & 4.87  & 1.22 &2.52 &4.69\\[2pt]
   \hline
$H_{\delta}F$&&&&&&\\[2pt] 
  \hline
  \verb'B1' & 0.57 & 0.42& 0.77 &0.16 &0.33 &0.03\\[2pt] 
  \verb'B2' &-0.38 & 0.2 & 0.04 & 0.13& 0.12&0.03\\[2pt]
  \verb'B3' & 0.9  & 2.74& 3.62 &0.7  &1.8  &3.11\\[2pt]
   \hline
$H_{\gamma}A$&&&&&&\\[2pt] 
 \hline
  \verb'B1' & 0.62 &-5.18 & -4.79 &0.18 & 1.63& 4.79\\[2pt]
  \verb'B2' & -0.39&-6.13 & -6.29 &0.14 &3.16 &15.81\\[2pt]
  \verb'B3' & 2.21 &  0.68&2.83   & 1.72&1.93 & 0.23\\[2pt]
   \hline
$H_{\gamma}F$&&&&&&\\[2pt] 
 \hline
  \verb'B1' & 0.74 &  -1.34& -0.88 &0.21 &0.23 &0.34\\[2pt] 
  \verb'B2' & -0.22&-1.64  & -1.73 &0.08 &0.95 &1.14\\[2pt]
  \verb'B3' & 1.11 &2.07   & 3.14  &0.86 &1.7  &1.76\\[2pt]
   \hline
$H_{\beta}$&&&&&&\\[2pt] 
 \hline
  \verb'B1' & 0.5   &1.78 & 2.09 &0.14 & 0.77 &0.57\\[2pt] 
  \verb'B2' & -0.24 &1.75 & 1.65 &0.09 & 0.79 &1.29\\[2pt]
  \verb'B3' & 0.64  &3.14 & 3.76 & 0.5 & 1.89 &3.93\\[2pt]
   \hline
$H_{\beta}G$&&&&&&\\[2pt] 
 \hline
  \verb'B1' & 0.44  & 2.03 & 2.31 & 0.13 &  0.87&0.72\\[2pt] 
  \verb'B2' & -0.16 & 2.02 & 1.95 &0.06  &0 .98 &1.72\\[2pt]
  \verb'B3' &  0.55 & 3.2  & 3.73 &0.43  &1.84  &4.09\\[2pt]
   \hline
$CN_{1}$&&&&&&\\[2pt] 
 \hline
  \verb'B1' & 0.07  & 0.05 & 0.09 & 0.02& 0.07 &0.01\\[2pt] 
  \verb'B2' & 0.05  &0.09  & 0.11 & 0.02& 0.07 & 0\\[2pt]
  \verb'B3' & -0.03 &-0.08 & -0.11& 0.02& 0.06 & 0\\[2pt]
   \hline
$CN_{2}$&&&&&&\\[2pt] 
 \hline
  \verb'B1' & 0.05  &0.06  & 0.09  &0.02 &0.04 & 0\\[2pt] 
  \verb'B2' & 0.07  & 0.13 &  0.16 &0.02 & 0.1 &0.01\\[2pt]
  \verb'B3' & -0.03 &-0.04 & -0.07 &0.02 &0.04 & 0\\[2pt]
   \hline
$M{\gamma}_1$&&&&&&\\[2pt] 
 \hline
  \verb'B1' &0.03 & 0.1  &0.11  &0.01 &0.04 & 0 \\[2pt] 
  \verb'B2' & 0.03& 0.13 &0.15  & 0.01& 0.08& 0.01 \\[2pt]
  \verb'B3' &0    &0.03  & 0.03 & 0   & 0.01& 0 \\[2pt]
 \hline
$M{\gamma}_2$&&&&&&\\[2pt] 
 \hline
  \verb'B1' & 0.03& 0.23& 0.25 &0.01 &0.09 &0.01  \\[2pt] 
  \verb'B2' &0.04 &0.28 &0.3   &0.02 &0.16 & 0.03 \\[2pt]
  \verb'B3' &-0.03&0.11 &0.08  &0.02 &0.03 &0  \\[2pt]
   \hline
   \hline
\end{tabular}
\label{table:4}
\end{table}

Now that we can see which classes of indices influence our GP, we can study in more detail their internal structure. Therefore we analyze the relations between each index and the NodeLength, with the same methodology used with the general galaxy properties but for each abundance index in our input sample. See Table \ref{table:4}, \ref{table:5} and \ref{table:6}.

\begin{table}
 \caption{{\bf Same as in table \ref{table:4}, but for a different set of indices.}}
 \begin{tabular}{lllllll}
 \hline
    Branch & Slope & $y_{intercept}$ & $\overline{X}_{OLS}$ & $\sigma_{OLS}$ & $\sigma_{WLS}$ & $X_{i}^{2}$  \\
 \hline
 \hline
$M{\gamma}_B$&&&&&&\\[2pt] 
 \hline
  \verb'B1' & 0.15 &3.81 &  3.9 & 0.04& 1.4&2.57  \\[2pt] 
  \verb'B2' & 0.6  &4.45 & 4.7  &0.21 &2.38 &8.34  \\[2pt]
  \verb'B3' &-0.03 &0.11 &  0.08&0.02 &0.03 & 0  \\[2pt]
   \hline
\verb"Ca4227"&&&&&&\\[2pt] 
 \hline
  \verb'B1' & -0.13& 1.21 & 1.13 &0.04 &0.4  & 0.26 \\[2pt] 
  \verb'B2' & -0.03& 1.28 & 1.27 &0.01 &0.76 & 0.73 \\[2pt]
  \verb'B3' &-0.11 & 0.59 &  0.48& 0.09&0.18 &0.14  \\[2pt]
   \hline
\verb"G4300"&&&&&&\\[2pt] 
 \hline
  \verb'B1' & -0.27& 5.1 & 4.93 &0.08 &1.91 &4.54  \\[2pt] 
  \verb'B2' &0.02  &5.53 & 5.54 &0.01 &2.98 &13.0  \\[2pt]
  \verb'B3' &-1.08 &2.25 & 1.2  &0.84 &0.2  &2  \\[2pt]
   \hline
\verb"Fe4383"&&&&&&\\[2pt] 
 \hline
  \verb'B1' &0.5   &4.23 & 4.54 &0.14 & 1.89&3.24  \\[2pt] 
  \verb'B2' &-0.01 & 5.2 & 5.2  & 0   & 3.26& 12.54 \\[2pt]
  \verb'B3' &-1.12 & 2.75& 1.67 &0.87 &0.29 &3.07  \\[2pt]
   \hline
\verb"Fe4531"&&&&&&\\[2pt] 
 \hline
  \verb'B1' &-0.28 &3.43  & 3.25 &0.08 &1.31 &2.09  \\[2pt] 
  \verb'B2' & 0.02 &3.52  & 3.53 &0.01 &1.84 &5.25  \\[2pt]
  \verb'B3' &-0.62 & 2.65 & 2.04 &0.49 &0.84 &2.76  \\[2pt]
   \hline
\verb"Fe4930"&&&&&&\\[2pt] 
 \hline
  \verb'B1' &0.31 & 1.73& 1.93 & 0.09 &0.84 & 0.57 \\[2pt] 
  \verb'B2' &0.11 &1.89 & 1.93 &0.04 & 1.08& 1.53 \\[2pt]
  \verb'B3' &-0.21& 1.31&1.11  &0.16 &0.29 & 0.76 \\[2pt]
   \hline
\verb"Fe5015"&&&&&&\\[2pt] 
 \hline
  \verb'B1' & 0.23& 4.46& 4.6  &0.06 &1.58 & 3.63 \\[2pt] 
  \verb'B2' & 0.19&5.13 & 5.22 &0.07 &2.68 & 11.11 \\[2pt]
  \verb'B3' &-0.74&3.28 &2.56  &0.57 &0.54 & 4.81 \\[2pt]
   \hline
\verb"Fe5270"&&&&&&\\[2pt] 
 \hline
  \verb'B1' &0.17  & 2.78&2.88  &0.05 &1.24 & 1.45 \\[2pt] 
  \verb'B2' &0     & 3.0 & 3.0  &0    & 1.71&3.92  \\[2pt]
  \verb'B3' & -0.46& 1.97& 1.52 & 0.36&0.44 & 1.58 \\[2pt]
   \hline
\verb"Fe5270S"&&&&&&\\[2pt] 
 \hline
  \verb'B1' & 0.03 &2.19 & 2.21 &0.01 &0.85 & 0.83  \\[2pt] 
  \verb'B2' &-0.03 &2.31 & 2.3  &0.01 & 1.3 & 2.32  \\[2pt]
  \verb'B3' & -0.36& 1.57& 1.22 &0.28 & 0.39& 1.0   \\[2pt]
\hline
  \verb"FE5335"&&&&&&\\[2pt] 
 \hline
  \verb'B1' &0.63 & 2.15& 2.54 & 0.18 & 1.19 & 1.02 \\[2pt] 
  \verb'B2' &0.1  &2.71 & 2.75 & 0.03 & 1.5  & 3.13\\[2pt]
  \verb'B3' &-0.38&1.62 & 1.25 & 0.3  & 0.39 & 1.06 \\[2pt]
   \hline
   \hline
\end{tabular}
\label{table:5}
\end{table}

If an absorption index has a good fit with NodeLength and, at the same time, has a different range of values for each GP, we conclude that this index is relevant for the hierarchical structure of our branches. 
\begin{table}

 \caption{ {\bf Same as in table \ref{table:4}, but for a different set of indices.}}
 \begin{tabular}{lllllll}
 \hline
    Branch & Slope & $y_{intercept}$ & $\overline{X}_{OLS}$ & $\sigma_{OLS}$ & $\sigma_{WLS}$ & $X_{i}^{2}$  \\
 \hline
 \hline
\verb"Fe5406"&&&&&&\\[2pt] 
 \hline
  \verb'B1' & 0.33&1.45 &1.66  &0.1  &0.79 & 0.49 \\[2pt] 
  \verb'B2' &0.03 & 1.8 & 1.82 &0.01 &0.99 &1.39  \\[2pt]
  \verb'B3' &-0.24& 1.06& 0.83 &0.19 &0.2 & 0.48 \\[2pt]
   \hline
\verb"FE5709"&&&&&&\\[2pt] 
 \hline
  \verb'B1' & -0.1 &0.93 & 0.86 & 0.03& 0.32& 0.15 \\[2pt] 
  \verb'B2' & -0.05& 0.87&  0.85& 0.02& 0.48& 0.33 \\[2pt]
  \verb'B3' &0.01  &0.47 & 0.48 &0.01 &0.25 & 0.09 \\[2pt]
  \hline
$OIII_{1}$&&&&&&\\[2pt] 
 \hline
  \verb'B1' &-0.13&-0.69  &-0.77   & 0.04& 0.29& 0.08 \\[2pt] 
  \verb'B2' & 0.07& -0.96 &  -0.93 &0.02 & 0.4 &0.39  \\[2pt]
  \verb'B3' & 0.13& -0.92 &-0.8    &0.1  &0.34 &0.34 \\[2pt]
   \hline
$OIII_{2}$&&&&&&\\[2pt] 
 \hline
  \verb'B1' & 0.11  & 0.97& 1.04 & 0.03 & 0.43 & 0.17  \\[2pt] 
  \verb'B2' & -0.02 &1.14 & 1.13 & 0.01 & 0.66 & 0.57  \\[2pt]
  \verb'B3' & -0.09 &0.38 & 0.29 & 0.07 & 0.63 & 1.5   \\[2pt]
   \hline
$Na_{D}$&&&&&&\\[2pt] 
 \hline
  \verb'B1' & 1.15 & 3.23 & 3.94 & 0.33 & 1.7  & 1.94  \\[2pt] 
  \verb'B2' & 1.06 & 4.42 & 4.86 & 0.38 & 2.45 & 8.25  \\[2pt]
  \verb'B3' & -0.36& 1.5  & 1.15 & 0.28 & 0.39 & 0.9   \\[2pt]
   \hline
\verb"C4668"&&&&&&\\[2pt] 
 \hline
  \verb'B1' & 0.23 & 6.98& 7.12 &0.06 &2.55 & 8.61  \\[2pt] 
  \verb'B2' & 0.9  &7.49 & 7.87 &0.32 &4.39 & 23.94 \\[2pt]
  \verb'B3' &-1.12 &3.0  & 1.92 &0.87 &0.33 & 3.72 \\[2pt]
   \hline
\verb"Fe5782"&&&&&&\\[2pt] 
 \hline
  \verb'B1' &0.29 &0.59 & 0.77 &0.08 & 0.31 &  0.06 \\[2pt] 
  \verb'B2' &0.02 &0.84 & 0.85 &0.01 &0.53  &  0.33  \\[2pt]
  \verb'B3' &-0.11&0.55 & 0.44 &0.09 &0.12  &  0.13 \\[2pt]
   \hline
\verb"Ca4455"&&&&&&\\[2pt] 
 \hline
  \verb'B1' & 0.21  & 1.06& 1.19 & 0.06 &0.52 &0.22  \\[2pt] 
  \verb'B2' & 0.08  &1.34 & 1.37 & 0.03 &0.74 &0.76  \\[2pt]
  \verb'B3' & -0.17 &0.7  & 0.53 & 0.13 &0.16 & 0.2 \\[2pt]
   \hline
$TIO_{1}$&&&&&&\\[2pt] 
 \hline
  \verb'B1' &0    & 0.04 & 0.04 & 0&0.01 & 0  \\[2pt] 
  \verb'B2' &0.01 & 0.04 & 0.04 & 0&0.01 & 0  \\[2pt]
  \verb'B3' &0    & 0.01 & 0.01 & 0&0.01 & 0 \\[2pt]
   \hline
$TIO_{2}$&&&&&&\\[2pt] 
 \hline
  \verb'B1' &0       &0.08  & 0.08 &0     & 0.03 &0  \\[2pt] 
  \verb'B2' & 0      &0.09  & 0.09 &0     & 0.04 &0 \\[2pt]
  \verb'B3' & -0.01  & 0.04 & 0.03 & 0.01 & 0.01 &0  \\[2pt]
   \hline
   \hline
\end{tabular}
\label{table:6}
\end{table}

Our main findings are:

    $\bullet$ Absorption Balmer lines: from table \ref{table:3} we see that only the B2 population remains in our tree. Balmer lines decrease in strength as stellar populations get older, which is reflected in their influence on populations B1 and B3. This reinforce that population B1 and B2 are different, although both belong to the red sequence of the Coma cluster. From Table \ref{table:4} we see that Balmer lines have a good fit for each galaxy population, however  $H_{\gamma F}$ is an index that have different range of values for each galaxy population independent of the possible error influence ($\sigma_{wls}$). Therefore this index is important for breaking the degeneracy between age and horizontal branch morphology (in agreement with \citealt{schiavon2004}).

    $\bullet$ Insensitive to $\alpha/Fe$ enhancement: this abundance indices are relevant for the B1 population. This is because, as seen in Table \ref{table:3} and figure \ref{fig8}-B, the B1 population is not present in our recalculated tree. Analyzing individual abundance indices, the $NaD$ index (see \ref{table:6}) shows a good fit with a different value for each GP, although values for population B2 can be influenced by errors in our data. This absorption index measures Carbon, Magnesium and Sodium, elements that are produced and released through stellar winds to the interstellar medium by massive stars. This tells us that galaxies in B1 have undergone a recent period of star formation, and this is the reason for showing up in our phylogenetic tree as a separated GP.
    
    $\bullet$ Negative Response to $\alpha/Fe$ enhancement: at first glance from Table \ref{table:3}, we see that the re-calculated tree has tree main GPs. But if we look at figure \ref{fig8}-A, we notice that the B3 population is missing its reddest galaxies, mostly of lenticular morphology. Nevertheless B1 population that also have mostly lenticular galaxies remains in our re-calculated tree. From Table \ref{table:5} and Table \ref{table:6} the abundance indices show a good fit for populations B1 and B3, although this results can be affected by the uncertainties in our sample. Populations B1 and B3 have different abundances of $Fe5270S$ index, in spite of the uncertainties. This index measures Carbon, Iron and Magnesium, elements that are produced in Type Ia supernovae. Therefore the hierarchical structure of S0 galaxies from B3 depends on the presence of Type Ia supernovae, unlike S0 galaxies in B1.

    $\bullet$ Positive Response to $\alpha/Fe$ enhancement: from figure \ref{fig8}-C) we see that the presence of these lines is not relevant for the identification of hierarchical structures in our GP, as we obtain the same galaxy populations without these absorption indices. 
    
    $\bullet$ Emission Line $[OIII] \alpha 5007$, with $[OIII_{1}]$ and $[OIII_{2}]$ (\citealt{Gonzalez_2005}): the re-calculated tree without this index shows a few galaxies from B2 and B3 and does not show B1 population.
    This reinforce that our GPs are different. These indices present a good fit (see Table \ref{table:6} ), but all the GPs have similar abundances. Therefore we can not evaluate their influence in the internal structures of our GPs. 

From this analysis we can infer that the minimum amount of indices needed to perform a successful and consistent phylogenetic study is 12 see Table \ref{table:4}, \ref{table:5} and \ref{table:6}, and those include the following: $H\delta_{A}$,$H\delta_{F}$, $H\gamma_{A}$ , $H\gamma_{F}$, $Fe4383$, $C4668$, $H_{\beta}$, $Fe5270$, $Fe5270S$, $Fe5709$, $Fe5782$, $NaD$. We have investigated the possibility of carrying out such an  analysis with fewer indices. The results showed inconsistencies in the sense that the MC trials with bootstrapping delivered different outcomes in terms of tree structures. Hence, we do not necessarily recommend applying such a phylogenetic method with indices other than those indicated above.

\subsubsection{Field}

The environment of the cosmic structures in which galaxies reside affects their evolution (\citealt{Kauffmann_03}, \citealt{Kauffmann_04}, \citealt{Baldry_2006}, \citealt{Peng_2010}). Therefore, in order to assess the differential effects between high and low density environments, we generate a phylogenetic tree with 438 field galaxies homogeneously distributed across the sky (see figure \ref{fig10}), at redshifts range of $0.034 - 0.054$. From our consensus tree, we obtain a main branch with 46 galaxies that appears 561 times from 1,000 tree samples, and several minor structures (220 branches with 1, 2, 3, or 4 nodes and 172 in single branches without nodes). We interpret the main branch as our main GP.

\begin{figure*}
 \includegraphics[width=17cm]{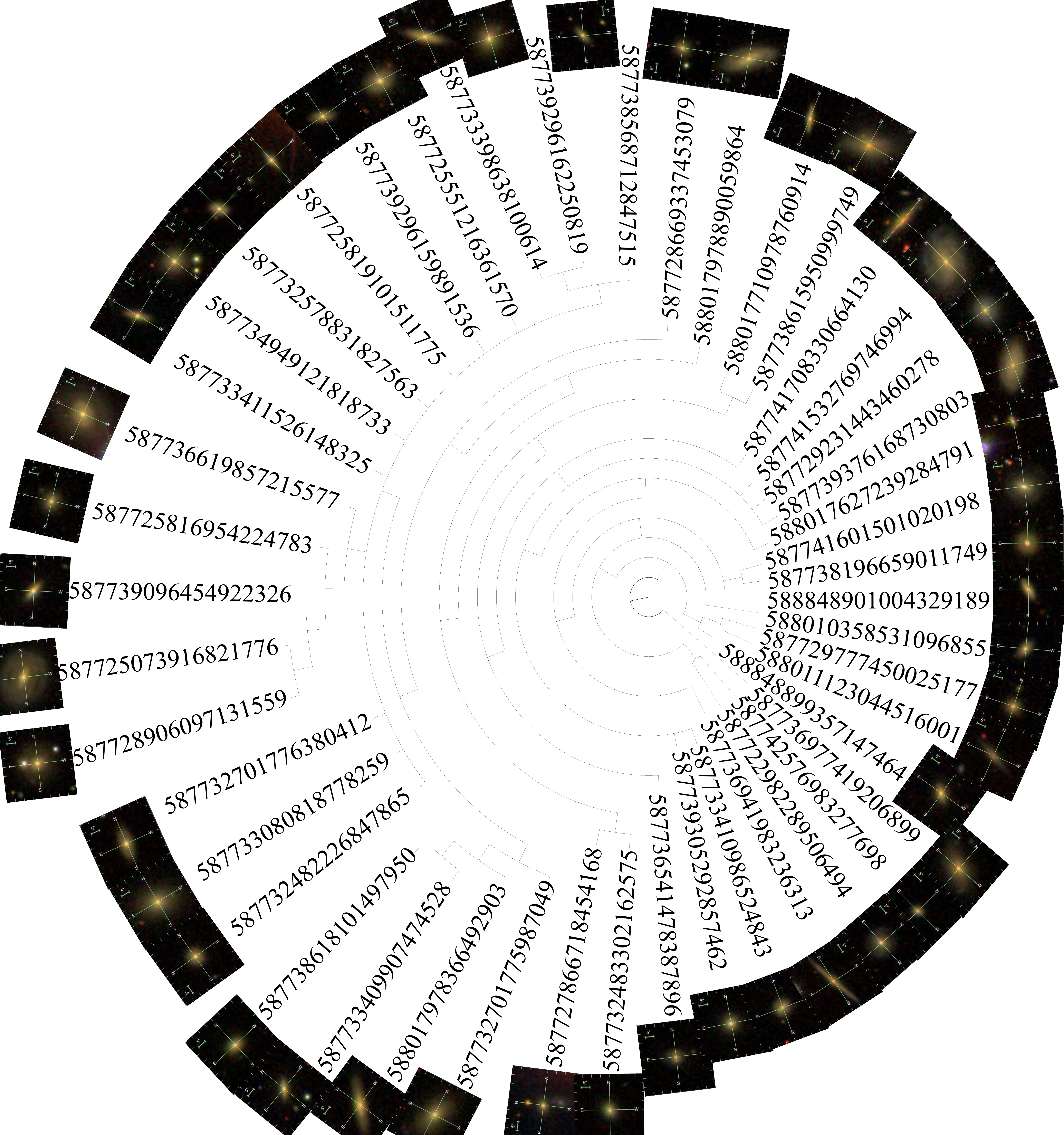}
 \caption{Galaxy population obtained in the field, plotted as a single tree. Each galaxy has its ID number from the SDSS DR7, and its image is from the SkyServer DR15.}
 \label{fig9}
\end{figure*}

\begin{figure*}
 \includegraphics[width=15cm, height= 17cm]{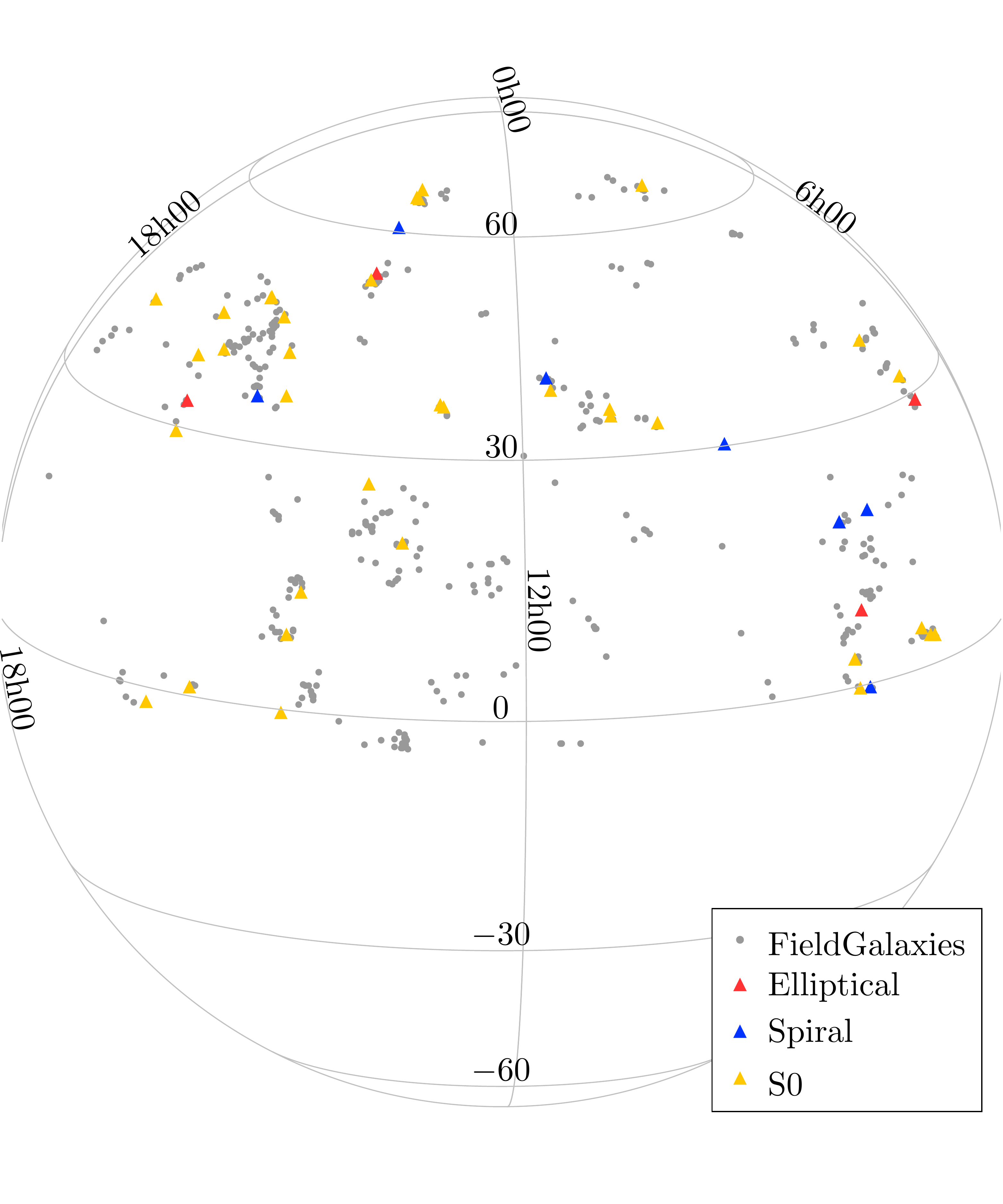}
 \caption{Spatial distribution of our field galaxy sample across the sky. Grey circles represent all galaxies in our sample, whereas triangles represent our main galaxy population. Red, yellow, and blue colors indicates, correspondingly, elliptical, S0 and spiral galaxies. Grey dots are galaxies from Coma that do not appear in the main GP.}
 \label{fig10}
\end{figure*} 

This population presents galaxies predominantly of lenticular morphology, that we can observe in more detail in Figure \ref{fig9}. Also show distributions characterized by high stellar masses  with a mean value of $log(M/M_{\odot}) = 10.63$ and 1-$\sigma$ dispersion of $log(M/M_{\odot}) = 0.31$. For the sSFR the distribution is centered at $10^{-10.4} yr^{-1}$, with 1-$\sigma$ dispersion of $0.42 yr^{-1}$, for the age the distribution is centered at $6.6 (Gyr)$  with 1-$\sigma$ dispersion of $3.52 Gyr$ and the distribution of the metallicity is centered at $[Z/H] =  0.01$ with 1-$\sigma$ dispersion of $0.23$. In order to analyze the internal structure of our field GP, we follow the same procedure as in section \ref{Coma}, comparing the lengths between nodes with common galaxy properties (see figure \ref{table:7}) with a bootstrap methodology generating 1000 fake data sets with replacement and performing a OLS fit. We also perform a WLS in order to assess the error in our absorption indices, setting the uncertainties equal to some intrinsic dispersion term $\sigma_int$ and the evaluating the value of this $\sigma_int$ term by forcing the reduced chi-squared of your fit to be unity. We report the value that gives $X_{i}^{2} = 1$, and the $\sigma_int$ for the WLS fit. 
\ref{table:7}.

\begin{table}
 \caption{Statistical Results between Nodelenght and galaxy properties for the field main population. We show the slope and y-intercept, with the 1-$\sigma$ uncertainties for a OLS fitting, as well as the 1-$\sigma$ dispersion for the WLS and the value for the chi squared forced to unity.}
 \begin{tabular}{lllllll}
  \hline
   Branch & Slope & $y_{intercept}$ & $\overline{X}_{OLS}$ & $\sigma_{OLS}$ & $\sigma_{WLS}$ & $X_{i}^{2}=1$  \\
  \hline
   \hline
\verb"Mass" &0.03  &10.53 & 10.63 &0.03 &3.27 &11.41\\[2pt] 
\verb"[Z/H]"&0.12  &-0.32 & 0.01  &0.11 &0.02 &0.01\\[2pt] 
\verb"Age"  & -0.63& 8.37 & 6.6   &0.6  &2.23 &8.03\\[2pt] 
\verb"SSFR" &-0.27 &-9.64 & -10.4 &0.26 &3.26 &9.6\\[2pt]
   \hline 
\end{tabular}
\label{table:7}
\end{table}

The internal structure of the main branch of our phylogenetic tree, shows a good fit with metallicity. In order to see whether a specific class of absorption indices influences the detection of this GP, we calculate a new phylogenetic tree, but excluding different classes of absorption lines, as we can see in figure \ref{fig12}. We notice that no particular class influences the outcome of our GP in the tree. 
We analyze the internal structure of our main GP, comparing the relation of the NodeLenght with respect to each absorption index. In order to know how the error in our data can affect our fit between absorption indices and the NodeLengthWe, we perform a OLS fit and a WLS fit with an intrinsic dispersion term to estimate its value by forcing the reduced chi-squared of our fit to be unity. We summarized the results in Table \ref{table:8}. The NodeLength have good fit for all the abundance indices. This shows that our GP is does not depend on any specific absorption line.

\begin{table}
 \caption{Statistical Results of OLS and WLS fit between Nodelenght and specific abundance lines for the main population of the Field sample.  We show the slope and y-intercept, with the 1-$\sigma$ uncertainties for a OLS fitting, as well as the 1-$\sigma$ dispersion for the WLS and the value for the chi squared forced to unity.}
 \begin{tabular}{lllllll}
  \hline
   Branch & Slope & $y_{intercept}$ & $\overline{X}_{OLS}$ & $\sigma_{OLS}$ & $\sigma_{WLS}$ & $X_{i}^{2}$  \\
  \hline
   \hline
$H_{\delta} A$ &-0.84 & 1.83 & -0.53 &0.8 &0.07 &0.65\\[2pt]
$H_{\delta} F$ & -0.33 &1.84 & 0.9  &0.32 &0.41 &0.61\\[2pt]
$H_{\gamma} A$ &-1.26&-1.19 & -4.75 & 1.2  &1.57 &0.14\\[2pt]
$H_{\gamma} F$ & -0.61&0.89 &-0.84  &0.58& 0.32&0.08\\[2pt]
$H_{\beta}$    &-0.08 &2.2 & 1.98 &0.07& 0.61&0.5\\[2pt] 
$H_{\beta}G $  &-0.14 &2.6 & 2.21 & 0.13& 0.64&0.7\\[2pt]
\hline
$CN_{1}$    &0.02& -0.05 &0.01&0.02& 0.01& 0\\[2pt] 
$CN_{2}$    &0&0.05 &0.06&0&0.03 &0 \\[2pt]
$MG_{1}$    &0.01& 0.05&0.08&0.01&0.03 &0\\[2pt]
$MG_{2}$    &0.02&0.15 &0.21&0.02& 0.06&0\\[2pt]
$MG_{B}$    &0.4& 2.28&3.41&0.38&1.04&0.54\\[2pt]
$Ca4227$    & 0.05& 0.77&0.92&0.05&0.17 &0.2\\[2pt] 
$G4300$     &0.57& 3.41&5.03&0.55& 1.68&1.28\\[2pt]
\hline
$Fe4383 $   &0.47 & 2.96 &4.28 &0.44&1.3 &0.91\\[2pt]
$Fe4531$    & 0.14&2.75 & 3.14&0.13&0.96 &0.78\\[2pt]
$Fe4930$    & 0&1.78 &1.78 &0& 0.46 &0.38\\[2pt]
$Fe5015$    &0.12 &4.41 &4.75 &0.12&1.3 &2.21\\[2pt]
$Fe5270$    & 0.26& 1.96&  2.68&0.25&0.83 &0.39\\[2pt]
$Fe5270S$   &0.1 & 1.77 &2.05 &0.1&0.61 &0.32\\[2pt]
$Fe5335$    &0.07 & 2.2 &2.41 &0.07& 0.6&0.68\\[2pt]
$Fe5406$    &0.06 & 1.38 &1.54 &0.05& 0.42&0.23\\[2pt]
$Fe5709 $   &-0.05 &1.02 &0.88 &0.05&0.29 &0.11\\[2pt]
\hline
$OIII_{1}$  & 0.05 &-0.75 & -0.6 & 0.05& 0.1& 0.12\\[2pt]
$OIII_{2}$  & 0.05& 0.99&1.13 & 0.05& 0.32 & 0.11\\[2pt]
\hline
$Ca4455 $   &0.14 &0.76 &1.15 &0.13 &0.39 &0.07\\[2pt]
$C4668$     &0.62 &3.56 &5.3 & 0.59&1.49 &1.6\\[2pt] 
$Fe5782$    &0.06 &0.5  & 0.68& 0.06&0.2  &0.03\\[2pt]
$Na_{D}$    & 0.32& 2.13& 3.03&0.3  &0.96 &0.47\\[2pt]
$TIO_{1}$   & 0   & 0.03&0.04 & 0   &0.01 & 0  \\[2pt]
$TIO_{2}$   & 0.01& 0.06& 0.07&0.01 &0.02 &0   \\[2pt]
   \hline 
\end{tabular}
\label{table:8}
\end{table}

\begin{figure*}
 \includegraphics[width=15cm, height= 15cm]{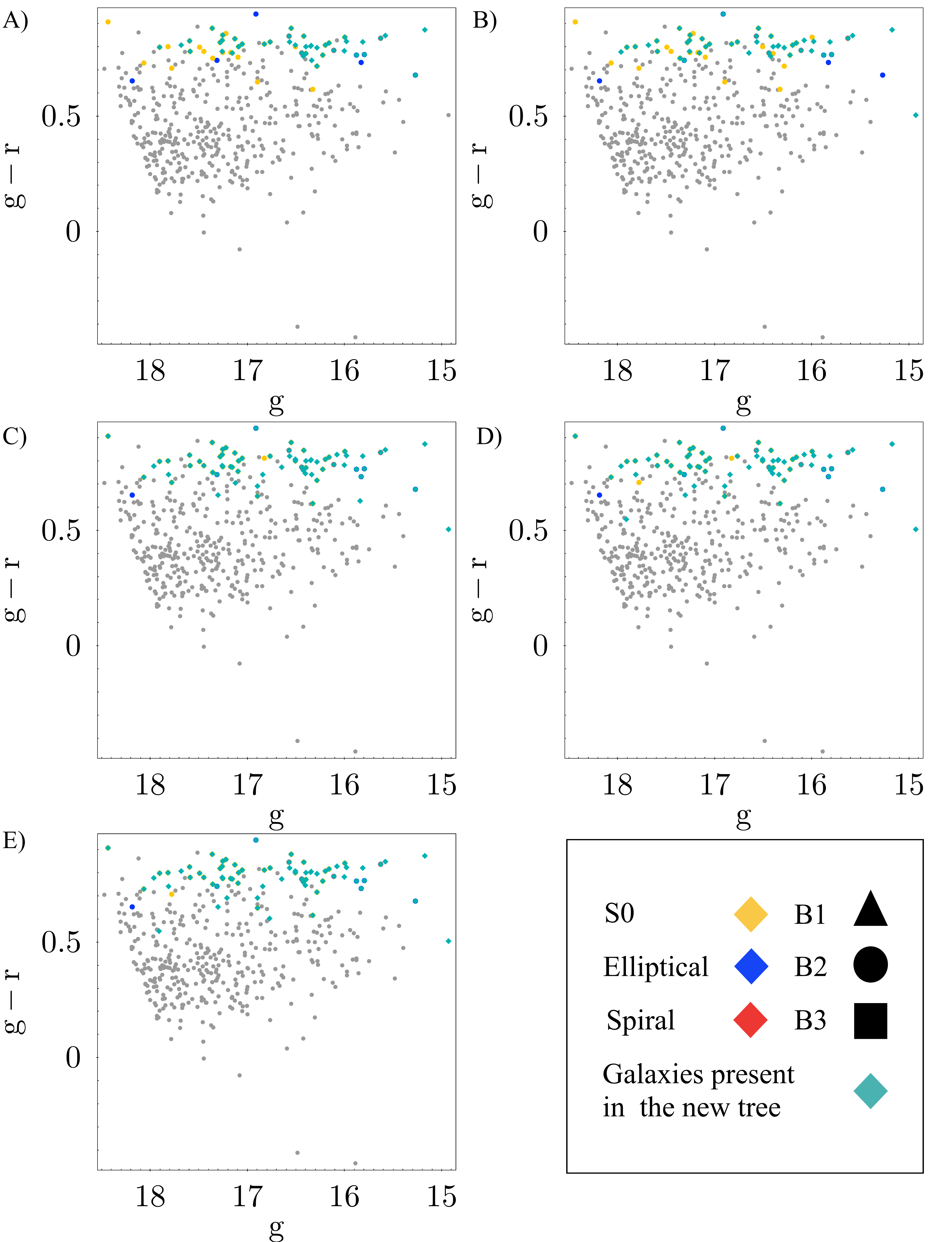}
 \caption{Color-magnitude diagram of our field galaxy sample, where panels A, B, C, D, and E show a recalculated phylogenetic tree excluding absorption Balmer lines, lines  insensitive to $\alpha/Fe$, lines with negative response to $\alpha/Fe$ enhancement, lines with positive response to $\alpha/Fe$ enhancement, and Emission Line $[OIII] \alpha 5007$, with $[OIII_{1}]$ and $[OIII_{2}]$ respectively. The cyan diamonds represent the galaxies that appear in branches of each recalculated tree. Grey dots represent galaxies from the field that do not appear ih the main GP.}
 \label{fig11}
\end{figure*}

\subsection{Discussion}

The phylogenetic analysis performed in this work takes as input information on the galaxy stellar population content in the form of a pairwise distance matrix whose components are the total difference between absorption indices for a given pair of galaxies. The NJ generates hierarchical structures that retrieve the evolutionary history of galaxies. The branches of the trees represent different GPs, and the NodeLength the chemical variation from one galaxy to another within the corresponding GP or branch. 
The phylogenetic analysis of galaxies is a better tool to identify GP because it allows us to detect them regardless of the morphology, stellar mass, sSFR, metallicity or stellar population properties of the galaxies.

\subsubsection{Comparison with other methods}

Chemical tagging is the closest method to our phylogenetic approach. Therefore, we make this in an unsupervised manner by performing a principal component analysis (PCA) of the same dataset of 30 abundance indices for the Coma Cluster and field samples.  Then we apply K-means clustering (\citealt{Hartigan_1979}), an unsupervised machine learning algorithm, to categorize our data into groups according to centroids. In order to obtain the centroids for the Coma Cluster we define 4 groups, so we can compare the result that we get with the phylogenetic approach. Then we also used the elbow method (\citealt{Joshi2013ModifiedKF}) to define the ideal number of groups, that in this case is 8 for both samples. 

\begin{figure*}
 \includegraphics[width=17cm, height=16cm]{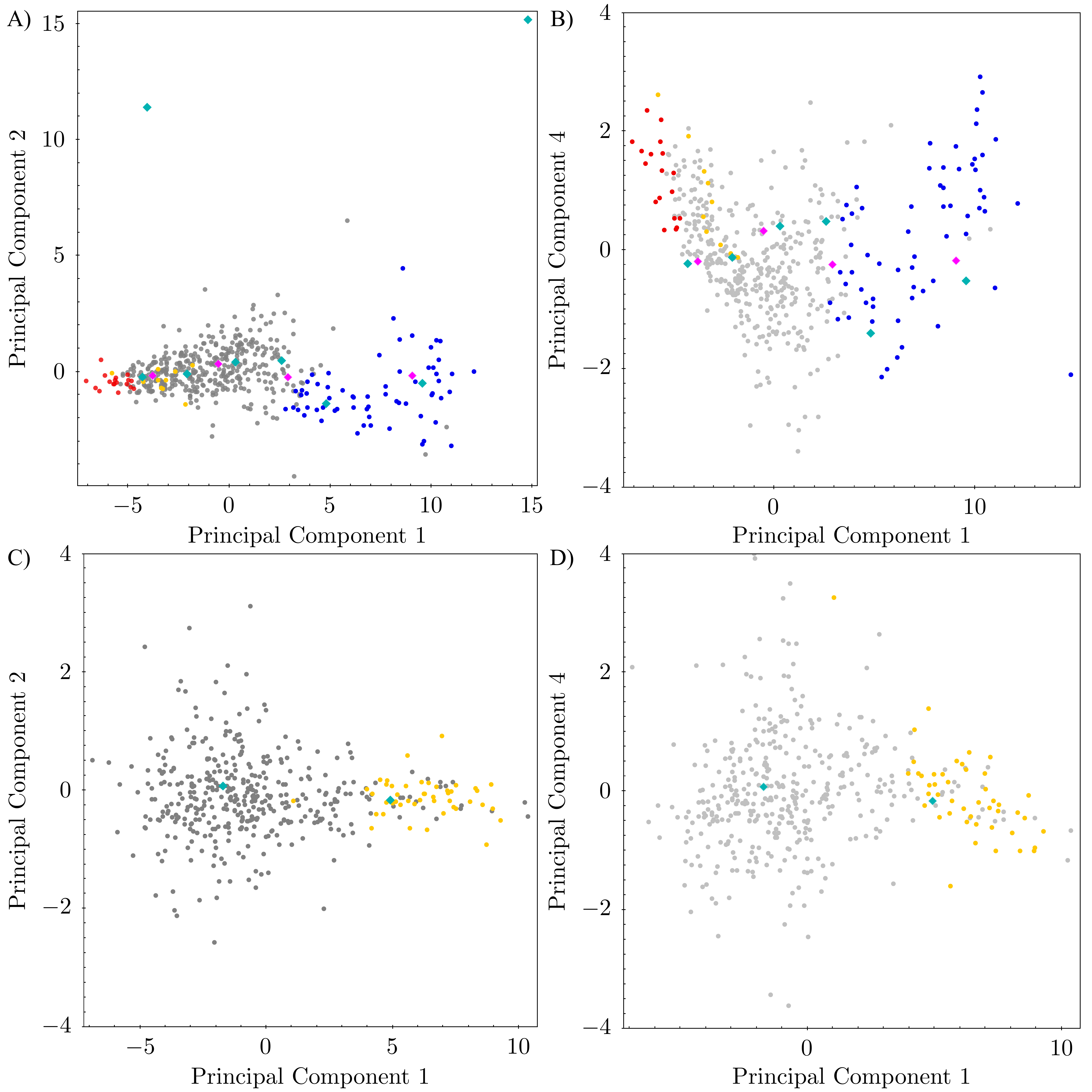}
 \caption{ We compare a PCA of 30 abundance indices with the phylogenetic approach, for our galaxy samples of the Coma Cluster and the field. Panel A presents the principal components 1 and 2, and panel B presents the principal components 1 and 4 for the Coma Cluster. Red, blue and yellow dots indicate the GP1, GP2 and GP3 of the Coma Cluster found by our consensus phylogenetic tree. Magenta and cyan diamonds indicate the 4 and 8 centroids for the PCA decomposition of the Coma cluster. 
 Panel C presents the principal components 1 and 2, and panel D presents the principal components 1 and 4 for the field. Yellow dots indicate the main GP found by our phylogenetic tree. Cyan diamonds indicate the 2 centroids for the PCA decomposition of the Field.}
 \label{fig12}
\end{figure*}

We calculate 10 principal components for the Coma Cluster and the field. Figure \ref{fig12} shows the combinations of principal components 1 and 2, and also principal components 1 and 4, which were the only combinations that showed class differentiation in both samples.

For the Coma Cluster, the PCA method at a first glance is able to separate two populations, with the combination of principal components 1 and 2. We calculate the centroids for 4 groups, and 8 groups based on the 10 components originally calculated. We see that the  centroids are not able to differentiate the three main GPs obtained with the phylogenetic approach. On the other hand, the combination of principal components 1 and 4 shows three possible groups. Nevertheless, by calculating the centroids for 4 and 8 groups we see that the centroids are unable to differentiate the GP obtained by our phylogenetic tree.

PCA decomposition is able at first glance to separate our sample into groups, but does not recover the GP obtained from the phylogenetic approach in the Coma Cluster. Furthermore, PCA lacks the extra parameter given by the phylogenetic approach that is the length between nodes. This new parameter can be used to explore correlations between specific absorption lines in order to understand the physical processes that are occurring in our GP. Also, the determination of groups depends on a K-means analysis which needs, as an input, a predefined number of groups in order to compute the centroids that separate our GP.

\subsubsection{Minor Structures}

Our phylogenetic analysis shows the presence of our GPs in both the Coma Cluster and the field. Nevertheless, for the Coma Cluster $30\%$ of our galaxies can be found in structures with one, two or three nodes, which we name minor structures, and $49\%$ in branches without nodes.  In a similar way, $50\%$ of field galaxy are found in minor structures and $39\%$ in branches without nodes.

For the Coma Cluster, the phylogenetic analysis shows three different GPs, represented by the three main branches of the tree. The NJ algorithm, in principle, is able to retrieve evolutionary information about their input taxa. Nevertheless, the input information can be incomplete, and thus the consensus tree restricts the possibility of further exploring structures with low statistical significance as our minor structures.

In order to explore the minor structures of our consensus phylogenetic tree, we divide them into branches without nodes, and  structures with 1, 2 or 3 nodes. We plot them in color-magnitude space and in the PCA analysis (see figure \ref{fig12}). From figure \ref{fig12}-A we see that these minor structures belong to the red sequence of the Coma Cluster. Nevertheless both samples of minor structures of the Coma Cluster do not show a clear disentanglement from each other. From figure \ref{fig12}-B the PCA decomposition is also unable to separate these minor structures as different groups. Therefore for the Coma Cluster we can conclude that the red sequence is not homogeneous, it has small groups of galaxies in color-magnitude space belonging to small substructures that appear in our consensus phylogenetic tree.  
For the field sample we make the same division for minor structures and from figure \ref{fig12} we can see that PCA decomposition is also unable to separate these minor structures in different groups. 

It is interesting to note that red sequence galaxies at intermediate and low stellar masses do not belong to any of the three main galaxy populations. This is interesting because at these masses most of the build-up of the red sequence should occur. The red sequence at intermediate and low stellar masses should in fact be a mixture of old galaxies and recently quenched galaxies. This probably reflects the difficulty in identifying more significant GP's. Furthermore, the signal-to-noise of the spectra is lower for faint galaxies with respect to bright ones, which may share part of the responsibility in not identifying definite structures. Larger samples of galaxies may help to resolve structures in the faint region of the red sequence.

\begin{figure*}
 \includegraphics[width=18cm, height= 6cm]{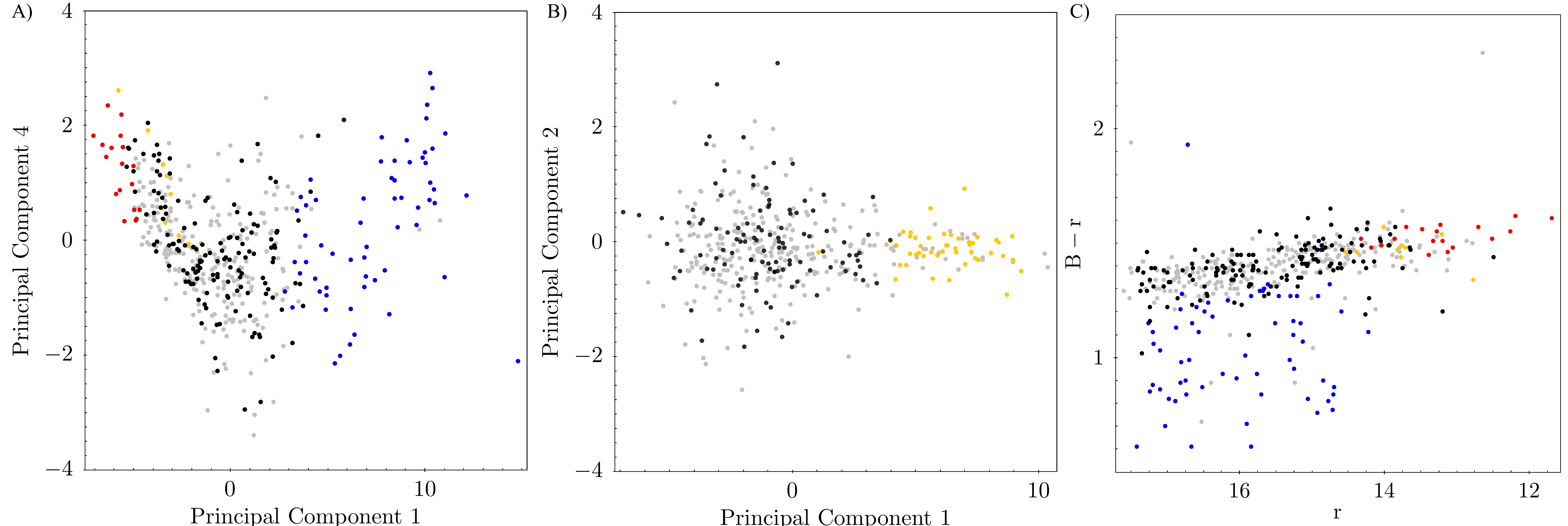}
 \caption{  Comparison of minor structures of the phylogenetic approach with the PCA. Panel A shows a PCA of the Coma Cluster. Red, blue and yellow dots represent GP1, GP2 and GP3 respectively. Panel B shows a PCA decomposition of the field. Yellow dots represent galaxies that belong to the main GP found in the field. Panel C shows a Color-Magnitude diagram of the Coma Cluster with the photometric filters Johnson B and Sloan r. For all Panels grey dots indicate galaxies that belong to structures with 1, 2 and 3 nodes and black dots represent galaxies in a single branch.}
 \label{fig13}
\end{figure*}

\section{Summary and Conclusions}
 
In this paper, we study the chemical properties of galaxies applying for the first time a phylogenetic techniques based on a distance matrix between absorption indices. This novel method gives us a graphical visualization of the chemical diversification among galaxies in the form of the hierarchical structure of a phylogenetic tree. 

This structure connects galaxies through features that we call "nodes". Branches in the consensus phylogenetic tree are interpreted as GPs, and the length between nodes (NodeLength) as the internal chemical variation along a branch. The  NodeLength acts as a chemical index, by relating galaxy stellar populations to various structural and environmental properties, both in galaxy groups or clusters and in the field.

We apply the phylogenetic method to 475 galaxies in the Coma cluster ($0.015 < z <0.033$) and 438 galaxies in the field ($0.035< z <0.054$) using 30 abundance indices from the Sloan Digital Sky Survey. 

The three main branches in the Coma Cluster are named B1, B2, and B3. Galaxies belonging to B1 are quiescent with high stellar mass with a mean value of $log(M/M_{\odot}) = 10.66$ and 1 $\sigma$ dispersion of $ log(M/M_{\odot}) = 0.21 $. Also this GP show high metallicity with a distribution centered at $[Z/H] = 0.38$ and 1-$\sigma$ dispersion of $ [Z/H] = 0.1)$. On the other hand this GP present a low sSFR with a distribution centered at a value of $10^{-10.75} [yr
^{-1}]$ and 1-$\sigma$ dispersion of $0.13 yr^{-1}$. This GP shows of predominant S0 morphology with a distribution of young-to-intermediate age stellar populations centered at $3.82 Gyr$ and 1-$\sigma$ dispersion of $1.15 Gyr$. On the other hand, B2 galaxies correspond to quiescent galaxies with high stellar mass with a mean value of $log(M/M_{\odot}) = 11.08$ and 1-$\sigma$ dispersion of  $log(M/M_{\odot}) = 0.33$. This GP has high metallicity with a distribution centered at $[Z/H] = 0.37$ and 1-$\sigma$ dispersion of $ [Z/H] = 0.07$), low sSFR with a mean value of $10^{-10.95} yr^{-1}$ and 1-$\sigma$ dispersion of $0.18 yr^{-1}$. B2 GP present intermediate-to-old-age stellar populations with a distribution centered at $ 9.0 Gyr $  with 1-$\sigma$ dispersion of  $2.85 Gyr$ and mostly elliptical morphology. B3 galaxies have low stellar mass with a mean value of $log(M/M_{\odot}) = 9.47$ and 1-$\sigma$ dispersion of $log(M/M_{\odot}) = 0.51$, with low metallicity distribution centered at $[Z/H] = -0.33$ and 1-$\sigma$ dispersion of $log(M/M_{\odot}) = 0.49$ and high sSFR centered at $10^{-9.64} yr^{-1}$ with 1-$\sigma$ dispersion of $1.8 yr^{-1}$. This GP have predominantly spiral morphology and contain stellar populations having a wide range of ages $2.62 Gyr$ with 1-$\sigma$ dispersion of $ 2.41 Gyr$. B2 is distributed in two clumps in the cluster, one in the centre including both BCGs NGC4874 and NGC4889 and the other near the galaxy NCG 4839 which is associated with a subgroup of galaxies falling into Coma (\citealt{Neumann_2001}).
Although the B1 and B2 populations reside in the red sequence they have different ranges of stellar ages and mass. Minor structures in the Coma cluster belong to the red sequence. Therefore, this one is not homogeneously distributed in color-magnitude space.

The metallicity plays an important role in the formation of the hierarchical structures of the different branches of our phylogenetic tree (see Table \ref{table:2}, Table \ref{table:7}).

B3 comprises a portion of galaxies with lenticular morphology (see figure \ref{fig8}) that are chemically different (negative response to $\alpha/Fe$ enhancement, Fe5270S) to lenticular galaxies in B1 (insensitive response to$\alpha/Fe$ enhancement, $Na_{D}$), suggesting that galaxies with lenticular morphology in Coma follow  different chemical paths. These can be interpreted as B1 lenticulars having undergone a recent period of star formation, while B3 lenticulars being abundant in type I SNe. 
\\

The galaxy population in the field is too homogeneous for us to be able to differentiate more than one population. The field presents a main branch with galaxies that we interpret as our main GP. This one has objects of predominantly lenticular morphology (see figure \ref{fig9}), with distributions characterized by high stellar masses with a mean value of $log(M/M_{\odot}) = 10.63$ and 1-$\sigma$ dispersion of $log(M/M_{\odot}) = 0.31$. For the sSFR the distribution is centered at $10^{-10.4} yr^{-1}$, with 1-$\sigma$ dispersion of $0.42 yr^{-1}$, for the age the distribution is centered at $6.6 Gyr)$ with 1-$\sigma$ dispersion of $3.52 Gyr$ and the distribution of the metallicity is centered at $[Z/H] =  0.01$) with $1-\sigma$ dispersion of $0.23$. The metallicity plays an important role in the hirarchical structure of the main populations (see table \ref{table:7}), as seen in the hirerarchical structures of the different GPs of the Coma cluster. The main population of the field shows no dependence of any specific absorption lines (see table \ref{table:8}), in contrast with what we observe in the GPs of the Coma cluster.     

Our phylogenetic approach introduces a new index that reflects the chemical distance between galaxies in a GP. This index is given by the NodeLength and allows us to study the hierarchical structures within the tree in terms of general properties of galaxies and specific abundance indices. 

From the analysis of the NodeLength values and specific absorption indices in both the field and the Coma Cluster, we find that  just 12 indices are relevant to find GPs with this phylogenetic approach (see Table \ref{table:4}), Table \ref{table:5} and Table \ref{table:6}, and those include the following: $H\delta_{A}$, $H\delta_{F}$, $H\gamma_{A}$ , $H\gamma_{F}$, $Fe4383$, $C4668$, $H_{\beta}$, $Fe5270$, $Fe5270S$, $Fe5709$, $Fe5782$, $NaD$.

As an overall conclusion of this work, we find that phylogenetics is a promising method to identify and study GPs in different environments. This is due to the fact that it is directly based on chemical information that is determined by the stellar population content in galaxies themselves. Our results indicate that, at least in the case of the Coma Cluster, galaxies in that high density region are chemically more heterogeneous than their counterparts in the field. 

We also notice that the hierarchical structure of the phylogenetic tree does not depend on the presence of positive response to $\alpha/Fe$ enhancement abundance indices. We speculate that these differences and the independence of $\alpha/Fe$ enhancement may be a consequence of environmental processes that drive the evolution of galaxies at different local densities. 

The main advantage of the phylogenetic approach is that, by being exclusively based on chemical information, it allows us to perform robust analyses and interpretations without introducing biases by first selecting samples based on a specific characteristic such as stellar mass, morphology, sSFR or metallicity, as done with more traditional techniques. We caution, however, that more studies with more cluster and field samples at higher redshifts are needed to firmly establish the validity of the results obtained in this work. The samples used in the phylogenetic analysis of galaxy spectra could suffer biases due to the spectroscopic selection function (completeness in flux, colour, morphological type, etc.).

\section*{Data Availability Statement}
There are no new data associated with this article. Please refer to the text for the references where the data were obtained from.
\section*{Acknowledgements}

The authors thank Inger J\o{}rgensen for useful discussions on the analyses of this work.

We acknowledge support by CONICYT Programa de Astronom\'ia Fondo ALMA-CONICYT 2017 31170002. R.D. gratefully acknowledges support from the Chilean Centro de Excelencia en Astrof\'isica y Tecnolog\'ias Afines (CATA) BASAL grant AFB-170002. P.C. acknowledges the support from the ALMA-CONICYT grant no 31180051.  NWCL gratefully acknowledges support from a Fondecyt Iniciacion grant 11180005. G.C.V. acknowledges support from CONICYT through the FONDECYT Initiation grant No. 11191130 and the Ministry of Economy, Development, and Tourism's Millennium Science Initiative through grant IC120009, awarded to The Millennium Institute of Astrophysics (MAS).




\bibliographystyle{mnras}
\bibliography{biblio} 








\bsp	
\label{lastpage}
\end{document}